\documentclass[aps,pre,twocolumn,groupedaddress,amsmath,amssymb,amsfonts]{revtex4-1}
\usepackage{txfonts}
\usepackage{bm} 
\usepackage{subfigure}
\usepackage{mathrsfs}
\usepackage{array}
\usepackage{amsmath,latexsym}
\usepackage{amsfonts} 
\usepackage{amssymb} 
\usepackage{graphicx, color}
\usepackage{algorithm}
\usepackage{algorithmic}
\begin{document}
\title{Dynamics of Order Parameters of  Non-stoquastic Hamiltonians in the Adaptive Quantum Monte Carlo Method}

\author{Shunta Arai$^1$}
\email[]{shunta.arai.s6@dc.tohoku.ac.jp}
\author{Masayuki Ohzeki$^{1,2}$}
\author{Kazuyuki Tanaka$^1$}
\affiliation{$^1$Graduate School of Information Sciences, Tohoku University, Sendai 980-8579, Japan\\
$^2$Institute of Innovative Research, Tokyo Institute of Technology, Nagatsuta-cho 4259, Midori-ku, Yokohama, Kanagawa, 226-8503 Japan}


\date{\today}

\begin{abstract}
We derive macroscopically deterministic flow equations with regard to the order parameters of the ferromagnetic $p$-spin model with infinite-range interactions. The $p$-spin model has a first-order phase transition for $p>2$. In the case of $p\geq5$ ,the $p$-spin model with anti-ferromagnetic XX interaction has a second-order phase transition in a certain region. In this case, however, the model becomes a non-stoqustic Hamiltonian, resulting in a negative sign problem. To simulate the $p$-spin model with anti-ferromagnetic XX interaction,  we utilize the adaptive quantum Monte Carlo method. By using this method, we can regard the effect of the anti-ferromagnetic XX interaction as  fluctuations of the transverse magnetic field.  A previous study (J.Inoue, J. Phys.Conf. Ser.233, 012010, 2010) derived deterministic flow equations of the order parameters in the quantum Monte Carlo method. In this study,  we derive macroscopically deterministic flow equations for the magnetization and transverse magnetization from the master equation in the adaptive quantum Monte Carlo method. Under the Suzuki--Trotter decomposition, we consider the Glauber-type stochastic process. We solve these differential equations by using the Runge--Kutta  method and verify that these results are consistent with the saddle-point solution of mean-field theory. Finally, we analyze the stability of the equilibrium solutions obtained by the differential equations. 
\end{abstract}
\pacs{}
\maketitle
\section{\label{sec:sec1}Introduction}
Finding the best solution in combinatorial optimization problems is computationally intractable.
Nevertheless, efficient solutions to such problems have been studied in various fields. 
Quantum annealing (QA) stochastically  solves combinatorial optimization problems  with the aid of  quantum fluctuations \cite{kadowaki_nishimori,qa2}.
To do so, the cost function is regarded as the physical energy of the system, and the minimizer  corresponds to the ground state of the physical system. 

The protocol of QA is realized in an actual quantum device using present-day technology, namely quantum annealer \cite{Dwave2010a,Dwave2010b,Dwave2010c,dwave_machine,Dwave2014,dwave3,dwave2}.
The output from the current version of the quantum annealer, D-Wave 2000Q, is not always minimizer due to limitation of the device and environmental effects \cite{Amin2015}.
Nevertheless the quantum annealer has been tested for numerous applications such as portfolio optimization \cite{Rosenberg2016}, protein folding \cite{Perdomo2012}, the molecular similarity problem \cite{Hernandez2017}, computational biology \cite{Richard2018}, job-shop scheduling \cite{Venturelli2015}, traffic optimization \cite{Neukart2017}, election forecasting \cite{Henderson2018}, machine learning \cite{Crawford2016,Neukart2018,Khoshaman2018}, and automated guided vehicles in plant \cite{Ohzeki2019}.
In addition, researches on implementing various problems into quantum annealer have been performed \cite{Arai2018nn,Takahashi2018,Ohzeki2018NOLTA,Okada2019}. 

In conventional  QA,  the Hamiltonian is described by 
$\mathscr{H} =\mathscr{H}_{0}+\mathscr{H}_{1}$.
The symbol $\mathscr{H}_{0}$ denotes the target Hamiltonian where we want to solve the ground state. This Hamiltonian consists of $z$ components of Pauli matrices $(\sigma_1^z,\cdots,\sigma_N^z)$, where  $N$ is the total number of spins. We add the quantum driver Hamiltonian $\mathscr{H}_{1}$ as the quantum fluctuation,
which is written $\mathscr{H}_{1}=  -\Gamma \sum_{i=1}^N \sigma_{i}^{x}$.
Here, $\Gamma$ denotes the strength of the transverse magnetic field, and $\sigma_{i}^{x}$ is the $x$ component of the Pauli matrix at site $i$. In QA, we initially set the strength of the transverse magnetic field very h igh such that the system explores a wide range of the state space to obtain the ground state. We then gradually decrease the strength of the transverse magnetic field. Following the Schr\"{o}dinger equation, the initial ground state adiabatically evolves into a nontrivial final ground state, which is the solution to the combinatorial optimization problem. 

The theoretically sufficient condition to obtain the ground state in QA is assured by the quantum adiabatic theorem \cite{adiabatic_therem}. The total  evolutionary  time  $\tau$ of  the Schr\"{o}dinger equation to obtain the ground state depends on the minimum energy gap $\Delta_{\rm{min}}$ from the ground state:
$\tau \gg \Delta_{\rm{min}}^{-2}$. In a system with a second-order phase transition, the minimum energy gap  polynomially decays with the system size as $\Delta_{\rm{min}}\propto N^{-\alpha}$. Thus, the  total evolutionary  time polynomially increases: $\tau \propto N^{2\alpha}$. In this case, QA efficiently solves the problems. On the other hand, when the system has a first-order phase transition, the minimum energy gap exponentially  decays with the system size as $\Delta_{\rm{min}}\propto \exp(-\alpha N)$. Since the total evolutionary time exponentially increases such that $\tau \propto \exp(2 \alpha N)$, QA cannot be performed efficiently. 

Seki and Nishimori proposed that QA with anti-ferromagnetic XX interaction can avoid the first-order phase transition for the ferromagnetic $p$-spin model with infinite-range interactions \cite{seki_antiferro,seki_hop,matsuura_prl,Okada_1dxx}. 
This entails an exponential efficiency speedup of conventional QA, because the second-order phase transition has a minimum energy gap that polynomially decreases as a function of the system size. 
A non-stoquastic Hamiltonian like the model with anti-ferromagnetic XX interaction cannot be simulated with the standard quantum Monte Carlo method (QMC) because the non-stoquastic Hamiltonian has positive values in off-diagonal elements in the computational basis to diagonalize the $z$ component of the Pauli matrix and has a negative sign problem \cite{suzu_toro,Bravyi2008,non_stoquastic,expo_enhancement}.

However, a method has been proposed to avoid the negative sign problem involved in  a particular class of non-stoquastic Hamiltonians \cite{adqmc}.
This method is called the adaptive quantum Monte Carlo method (AQMC). The AQMC treats the effect of the anti-ferromagnetic XX interaction as the fluctuation of the transverse magnetic field by using the delta function and its  Fourier integral transformation. We can calculate various physical quantities of the non-stoquastic Hamiltonian by estimating the transverse magnetization and changing the corresponding transverse magnetic field obtained by a saddle-point  solution. 

In this paper, we focus on  the dynamics of the AQMC. To simulate QA, we often utilize the QMC, which is mainly designed for sampling from a Boltzmann distribution. Although the dynamics of QMC differ from those of QA \cite{can_qmc_simulate_QA}, it has been found that some aspects  of the dynamics of QA can be expressed by QMC \cite{qmc_tunneling,scaling_qmc}. Therefore it is useful to consider the dynamics of QMC or AQMC for QA with and without a non-stoquastic Hamiltonian.  

We analyze the dynamics of the order parameters of $p$-spin model with anti-ferromagnetic XX interaction \cite{p_spin_kirk,glass_prb,Seoane,QAC,ensemble_pspin} . For cases with a stoquastic Hamiltonian , Inoue analytically derived macroscopically deterministic flow equations of the order parameter, for example longitudinal magnetization in infinite-range quantum spin systems \cite{dy_maxmum_like,determ_order_qmc,inoue_hop}.
The differential equations with respect to  the  macroscopic order parameter are obtained from the master equation  by considering the transition probability of the Glauber-type dynamics of microscopic states under the Suzuki--Trotter decomposition. 

Following this approach, we derive the macroscopically deterministic flow equations  of order parameters with  anti-ferromagnetic XX interaction in AQMC. The adaptive transverse magnetic field is changed by the transverse magnetization in AQMC. Therefore we newly introduce the dynamics of transverse magnetization and derive differential equations for magnetization and transverse magnetization. We compare the non-trivial behavior of the dynamics of order parameters with and without anti-ferromagnetic XX interaction.  

It is useful to establish a way of simulating  a class of non-stoquastic Hamiltonians with a classical computer in order to validate a new quantum annealer. To date, conventional QA with a transverse magnetic field has been implemented in the D-Wave machine.
However, a quantum annealer for non-stoquastic Hamiltonians is being developed. 
Analyzing the dynamics of order parameters with non-stoquastic Hamiltonians will help us to verify  the performance of this new quantum annealer for non-stoquastic Hamiltonians.
 
The remainder of this paper is organized as follows. 
In Sec. \ref{sec:sec2}, we show the algorithm for AQMC. 
In Sec. \ref{sec:sec3}, we derive the macroscopically deterministic flow equations with respect to a non-stoquastic Hamiltonian from the master equation. 
In Sec. \ref{sec:sec4}, we analyze the stability of the solutions obtained by the macroscopically deterministic flow equations.
In Sec. \ref{sec:sec5}, we show the numerical results of the differential equations of order parameters. 
Finally, in Sec. \ref{sec:sec6}, we summarize our results and discuss future research directions.

\section{\label{sec:sec2}Adaptive Quantum Monte Carlo Method}
In this paper, we treat the ferromagnetic $p$-spin model with infinite-range interactions as the target Hamiltonian which is written as 
\begin{equation}
\mathscr{H}_{0}=  -N\left(\frac{1}{N}\sum_{i=1}^{N} \sigma_{i}^{z} \right)^p.
\end{equation}
We add the quantum driver Hamiltonian as 
\begin{equation}
\mathscr{H}_{1} =-\Gamma \sum_{i=1}^N \sigma_{i}^{x}+\frac{\gamma}{2N}\left(\sum_{i=1}^N \sigma_{i}^{x}\right)^2,
\end{equation}
where $\gamma$ is the strength of the XX interaction. 
The partition function of the total Hamiltonian is given by
\begin{align} 
&Z=\mathrm{Tr} \Biggl\{ \exp \left( N \beta \left(\frac{1}{N}\sum_{i=1}^{N} \sigma_{i}^{z} \right)^p \right. \nonumber \\
&+\left. \beta \Gamma \sum_{i=1}^N \sigma_i^x-N\frac{\beta \gamma}{2}\left(\frac{1}{N}\sum_{i=1}^{N}\sigma_i^x\right)^2\right)\Biggl\},
\end{align}
where $\beta$ is the inverse temperature.
 We employ the Suzuki--Trotter decomposition to divide the total Hamiltonian into two parts \cite{suzu_toro}. After that, we introduce the delta function as 
\begin{align}
 1=\int  dm_{xk} \delta\left(Nm_{xk}-\sum_{i=1}^N \sigma_{ik}^x\right).
 \end{align}
 We utilize the Fourier integral representation of the delta function and introduce the auxiliary variable $\tilde{m}_{xk}$ on the Trotter slice $k$. We obtain the partition function as 
\begin{align} 
&Z \approx \lim_{M \rightarrow \infty}\mathrm{Tr}\Biggl\{\prod_{k=1}^M \int dm_{xk}\int d\tilde{m}_{xk}\exp \left(\frac{N\beta}{M} \left(\frac{1}{N}\sum_{i=1}^{N} \sigma_{ik}^{z} \right)^p\right)\nonumber \\
&\times \exp \left(N\frac{\beta \Gamma}{M} m_{xk}-N\frac{\beta \gamma}{2M}m_{xk}^2-\frac{\beta}{M}\tilde{m}_{xk}\left(Nm_{xk}-\sum_{i=1}^N \sigma_{ik}^x\right)\right)\Biggl\},
\label{intef_part}
\end{align}
where $M$ is the Trotter number. We have dropped a trivial coefficient $1/2\pi$ in the above expression. 
This partition function (\ref{intef_part}) is equivalent to the partition function of the transverse field Ising model. Furthermore, we use static approximation $\tilde{m}_{xk}=\tilde{m}_x$  and $m_{xk}=m_x$ to simplify the problem. Finally, the partition function is written as  
\begin{align}
Z\approx&\ \lim_{M \rightarrow \infty}\sum_{\bm{\sigma}}\Biggl\{ \int dm_{x}\int d\tilde{m}_{x}\lambda^{NM} \phi^M(m_x,\tilde{m}_x)\exp\left(-H_{\mathrm{eff}}\right)
\Biggl\}, \label{intef_part_d}
\end{align}
where we define 
\begin{align}
&\lambda =\cosh \left( \frac{\beta}{M}\tilde{m}_x\right) \exp\left( -B \right), \\
&\phi(m_x,\tilde{m}_x)=\exp\left(N\frac{\beta \Gamma}{M} m_{x}-N\frac{\beta \gamma}{2M}m_{x}^2-N\frac{\beta}{M}\tilde{m}_{x}m_{x}\right),\\
&H_{\mathrm{eff}}=-\frac{N}{M} \sum_{k=1}^M\left(\frac{1}{N}\sum_{i=1}^{N} \sigma_{ik} \right)^p -\frac{B}{\beta}\sum_{i=1}^N \sum_{k=1}^M\sigma_{ik} \sigma_{ik+1},\label{eff_H}\\
&B=-\frac{1}{2}\log\tanh\left(\frac{\beta}{M}\tilde{m}_x\right).
\end{align}
Here, we regard $\sigma_{ik}^z$ as the classical spin $\sigma_{ik}\in\{-1,+1\}$. We rewrite  the $\mathrm{Tr}$ into the summation of classical spins as $\sum_{\bm{\sigma}}$.

In the thermodynamic limit, we may take the saddle-point in the integral. The saddle-point is evaluated by $\tilde{m}_x=\Gamma-\gamma m_x$. 
Thus, the instantaneous  transverse magnetic field is determined 
by the transverse magnetization $m_x$. 
To estimate the transverse magnetization, we consider the conditional probability distribution as
$P(\bm{\sigma}|\tilde{m}_x)=Z(\tilde{m}_x)^{-1}\exp\left(-\beta H_{\mathrm{eff}}\right)$
where $Z(\tilde{m}_x)=\sum_{\bm{\sigma}}\exp\left(-\beta H_{\mathrm{eff}}\right)$ is the partition function of the effective spin model defined by the  conditional probability distribution.
The transverse magnetization is written as
$m_x=\langle (NM)^{-1}\sum_{i=1}^N\sum_{k=1}^M \tanh\left(\beta \tilde{m}_x/M\right)^{\sigma_{ik}\sigma_{ik+1}}\rangle$
where the bracket denotes the expectation with respect to the weight of the conditional probability distribution $P(\bm{\sigma}|\tilde{m}_x)$. 
We can realize the classical simulation of  the non-stoquastic Hamiltonian with anti-ferro magnetic infinite-range XX interactions by estimating the transverse magnetization in the standard QMC method  
and updating the instantaneous  transverse magnetic field $\tilde{m}_x$ according to the saddle-point solution. 

\section{\label{sec:sec3}The Dynamics of Adaptive Quantum Monte Carlo Method}
In this section, we introduce the dynamics of the $p$-spin model with XX interaction in AQMC. 
Following Sec. \ref{sec:sec2}, we can rewrite the $p$-spin model with XX interaction to the $p$-spin model with the transverse magnetic field fluctuated by the transverse magnetization as Eq. (\ref{eff_H}). The fluctuated transverse magnetic field is determined by the saddle-point solution $\tilde{m}_x=\Gamma - \gamma m_x$. Here, the transverse magnetization is fixed by the last one before. 

The effective Hamiltonian (\ref{eff_H}) is a classical system under the Suzuki--Trotter decomposition. 
Therefore, the  dynamics of AQMC can be written as a Glauber-type stochastic process whose transition probability is given by 
\begin{align}
w_i(\bm{\sigma}_{k})=\frac{1}{2}[1-\sigma_{ik}\tanh\left(\beta\Phi_i(\bm{\sigma}_k:\sigma_{ik\pm1})\right)],
\label{trans_rate}
\end{align} 
where $ \Phi_i(\bm{\sigma}_k:\sigma_{ik\pm1})$ is  an  instantaneous  local field at site $i$ on the $k$-th Trotter slice as
\begin{align}
\beta \Phi_i(\bm{\sigma}_k:\sigma_{ik\pm1})=\frac{\beta p}{M} \left(\frac{1}{N}\sum_{i=1}^{N} \sigma_{ik} \right)^{p-1}+\frac{B}{2}\left(\sigma_{ik-1}+\sigma_{ik+1}\right).\label{local_cite}
\end{align}
Here, we neglect the term less than $O(1)$.

The master equation for the probability $p_{t}(\left\{\bm{\sigma}_k\right\})$, which represents the probability of a macroscopic state including the $M$-Trotter slices $\left\{\bm{\sigma}_k\right\}\equiv \left(\bm{\sigma}_1,\cdots,\bm{\sigma}_{M}\right)$, $\bm{\sigma}_{k}=(\sigma_{1k},\cdots,\sigma_{Nk})$ at time $t$  is written as follows:
\begin{align}
\frac{dp_{t}(\left\{\bm{\sigma}_k\right\})}{dt}=\sum_{k=1}^M\sum_{i=1}^N\left[ p_t(F_i^{(k)}(\bm{\sigma}_k))w_i(F_i^{(k)}(\bm{\sigma}_k)) -p_t(\bm{\sigma}_k)w_i(\bm{\sigma}_{k})\right],
\label{mastereq}
\end{align}
where we define the probability of a macroscopic state on the $k$-th Trotter slice as $p_t(\bm{\sigma}_k)$ and a spin flip operator $F_i^{(k)}(\bm{\sigma}_k)$  as 
\begin{align}
F_i^{(k)}(\bm{\sigma}_k)=(\sigma_{1k},\sigma_{2k},...,-\sigma_{ik},...,\sigma_{Nk}).
\end{align}
We impose the periodic boundary conditions $\bm{\sigma}_{1}=\bm{\sigma}_{M+1}$.
Next, we introduce a probability distribution $P_{t}(\left\{m_k\right\},\left\{m_{xk}\right\})$ of the macroscopic order parameters 
\begin{align}
m_k(\bm{\sigma}_k)&=\frac{1}{N}\sum_{i=1}^N\sigma_{ik},\\
m_{xk}(\bm{\sigma}_k,\bm{\sigma}_{k+1})&=\frac{K}{N}\sum_{i=1}^N\sigma_{ik}\sigma_{ik+1}
\end{align}
as 
\begin{align}
&P_{t}(\left\{m_k\right\},\left\{m_{xk}\right\})\nonumber \\
&=\prod_{k=1}^M\sum_{\bm{\sigma}_k}p_{t}(\bm{\sigma}_k)\delta(m_k-m_k(\bm{\sigma}_k))\delta(m_{xk}-m_{xk}(\bm{\sigma}_k,\bm{\sigma}_{k+1})),
\label{order_p}
\end{align}
where we define $K=(\tanh^2(\beta \tilde{m}_x/M)-1)/(2\tanh(\beta \tilde{m}_x/M))$ which is the correction term of the instantaneous transverse magnetic field generated from the derivative of $\log Z(\tilde{m}_x)$ , the set of the longitudinal magnetization as $\left\{m_k\right\}\equiv (m_1,\cdots,m_M)$, and the set of the transverse magnetization as $\left\{m_{xk}\right\}\equiv(m_{x1},\cdots,m_{xM})$. The notation $\sum_{\bm{\sigma}_k}$ is written as 
\begin{align}
\sum_{\bm{\sigma}_k}\left(\cdots\right)\equiv \sum_{i=1}^N\sum_{\sigma_{ik}=\pm1}\left(\cdots\right).
\end{align}
We regard $m_k$ as the magnetization on the Trotter slice $k$, and $m_{xk}$ as the transverse magnetization between the Trotter slice $k$ and $k+1$.
Following Ref. \cite{determ_order_qmc} , we take the derivative of Eq. (\ref{order_p}) with respect to $t$ and obtain differential equations as
\begin{align}
&\frac{dP_{t}(\left\{m_k\right\},\left\{m_{xk}\right\})}{dt}=\sum_{k} \frac{\partial  }{\partial m_k}m_k P_{t}(\left\{m_k\right\},\left\{m_{xk}\right\})\nonumber \\
&- \sum_{k}\frac{\partial  }{\partial m_k}\left\{P_{t}(\left\{m_k\right\},\left\{m_{xk}\right\})\frac{1}{N}\sum_{i=1}^{N}\tanh\left(\beta\Phi_i(\bm{\sigma}_k:\sigma_{ik\pm1})\right)\right\} \nonumber \\
&+\sum_{k}\frac{\partial  }{\partial m_{xk}}m_{xk} P_{t}(\left\{m_k\right\},\left\{m_{xk}\right\})\nonumber \\
&-\sum_{k} \frac{\partial  }{\partial m_{xk}}\left\{P_{t}(\left\{m_k\right\},\left\{m_{xk}\right\})\frac{K}{N}\sum_{i=1}^N\sigma_{ik+1}\tanh(\beta\Phi_i(\bm{\sigma}_k:\sigma_{ik\pm1}))\right\} .\label{order_mastert}
\end{align} 

From Eq. (\ref{order_mastert}), we can obtain these deterministic flow equations as 
\begin{align}
&\frac{dm}{dt}=\nonumber \\
&-m+\left(p m^{p-1}\right)\frac{\tanh\left(\beta \sqrt{(p m^{p-1})^2+(\Gamma-\gamma m_x)^2}\right)}{\sqrt{(p m^{p-1})^2+(\Gamma-\gamma m_x)^2}},
\label{dy_mz}
\end{align}
\begin{align}
&\frac{dm_x}{dt}=\nonumber \\
&-m_x+\left(\Gamma-\gamma m_x\right)\frac{\tanh\left(\beta \sqrt{(p m^{p-1})^2+(\Gamma-\gamma m_x)^2}\right)}{\sqrt{(p m^{p-1})^2+(\Gamma-\gamma m_x)^2}}.
\label{dy_mx}
\end{align}

Here, we use the saddle-point solution $\tilde{m}_x=\Gamma-\gamma m_x$.
The derivations of Eqs. (\ref{dy_mz}) and (\ref{dy_mx}) are described in Appendix A.

\section{\label{sec:sec4}Stability Analysis of The Equilibrium Solutions}
In this section, we derive the stability of the solutions obtained by the deterministic flow equations \cite{Strogatz,stability_1,stability_2}. To simplify the problem, we consider the zero-temperature limit $\beta \rightarrow \infty$. 
We can rewrite the deterministic flow equations  (\ref{dy_mz}) and (\ref{dy_mx}) as 
\begin{align}
\frac{dm}{dt}&=-m+\frac{pm^{p-1}}{\sqrt{(pm^{p-1})^2+(\Gamma-\gamma m_x)^2}}=f(m,m_x)\label{eq24},\\
\frac{dm_x}{dt}&=-m_x+\frac{\Gamma-\gamma m_x}{\sqrt{(pm^{p-1})^2+(\Gamma-\gamma m_x)^2}}=g(m,m_x)
\label{eq25}.
\end{align}
We assume the existence of the equilibrium solutions $(m^*,m_x^*)$. These solutions satisfy $f(m^*,m_x^*)=0$ and $g(m^*,m_x^*)=0$. We consider infinitesimal increments of  $m$ and $m_x$ around the equilibrium solutions as
 \begin{align}
m&=m^*+u\label{eq26},\\
m_x&=m_x^*+v\label{eq27}.
\end{align}
The Taylor expansions for $f(m, m_x)$ and $g(m, m_x)$ around the equilibrium solutions yield
\begin{align}
&f(m,m_x)\nonumber\\
&\approx f(m^*,m_x^*)+\left. \frac{\partial f }{\partial m}\right|_{m^*,m_x^*}(m-m^*)+\left. \frac{\partial f }{\partial m_x}\right|_{m^*,m_x^*}(m_x-m_x^*)\label{eq28},
\end{align}
and
\begin{align}
&g(m,m_x)\nonumber\\
&\approx g(m^*,m_x^*)+\left. \frac{\partial g }{\partial m}\right|_{m^*,m_x^*}(m-m^*)+\left. \frac{\partial g }{\partial m_x}\right|_{m^*,m_x^*}(m_x-m_x^*)\label{eq29}.
\end{align}
From Eqs. (\ref{eq26})--(\ref{eq29}), the time differential of infinitesimal increments are written as follows:
\begin{align}
\frac{d}{dt}
\left(
    \begin{array}{c}
     u\\
     v\\
       \end{array}
  \right)=\left(
    \begin{array}{cc}
      f_m & f_{m_x} \\
  g_m & g_{m_x}\\
       \end{array}
  \right)\left( \begin{array}{c}
     u \\
    v\\
       \end{array}\right),
\end{align}
where we define the Jacobian matrix as
\begin{align}
J\equiv \left(\begin{array}{cc}
      f_m & f_{m_x} \\
  g_m & g_{m_x}\\
       \end{array}
  \right).
\end{align}
To determine the stability of the equilibrium solutions, we consider the characteristic polynomial of the Jacobian matrix whose  eigenvalues are $\lambda_1$ and $\lambda_2$. By evaluating the trace of the Jacobian matrix $\mathrm{tr} J=\lambda_1+\lambda_2$ , the determinant $\mathrm{det} J=\lambda_1\cdot \lambda_2$, and whether the eigenvalues  are complex, we can investigate the stability of the equilibrium solutions. The condition with  real roots of eigenvalues is $(\mathrm{tr} J)^2>4\cdot \mathrm{det} J$. This condition determines whether the equilibrium solutions have  oscillations. 

\section{\label{sec:sec5}Experimental results}
In this section, we describe our numerical experiments. We numerically solve differential equations (\ref{dy_mz}) and (\ref{dy_mx}) using the Runge--Kutta  method with a sufficiently low-temperature $T$ and inverse temperature $\beta=1/T=100$.  The ferromagnetic $p$-spin model has a first-order phase transition for $p>2$ \cite{p_ising}. 
We set  the parameters $\gamma=0$ or $\gamma=18$. 
In the case of $\gamma=18$, the $p$-spin model has a second-order phase transition for $p>4$ following the phase diagram in \cite{seki_antiferro}. In this study, we consider $p=5$ and compare the dynamics of order parameters with  and without anti-ferromagnetic XX interaction. 
\begin{figure*}[t]
\subfigure[\label{fig:fig_1a}]{\includegraphics[width=50mm]{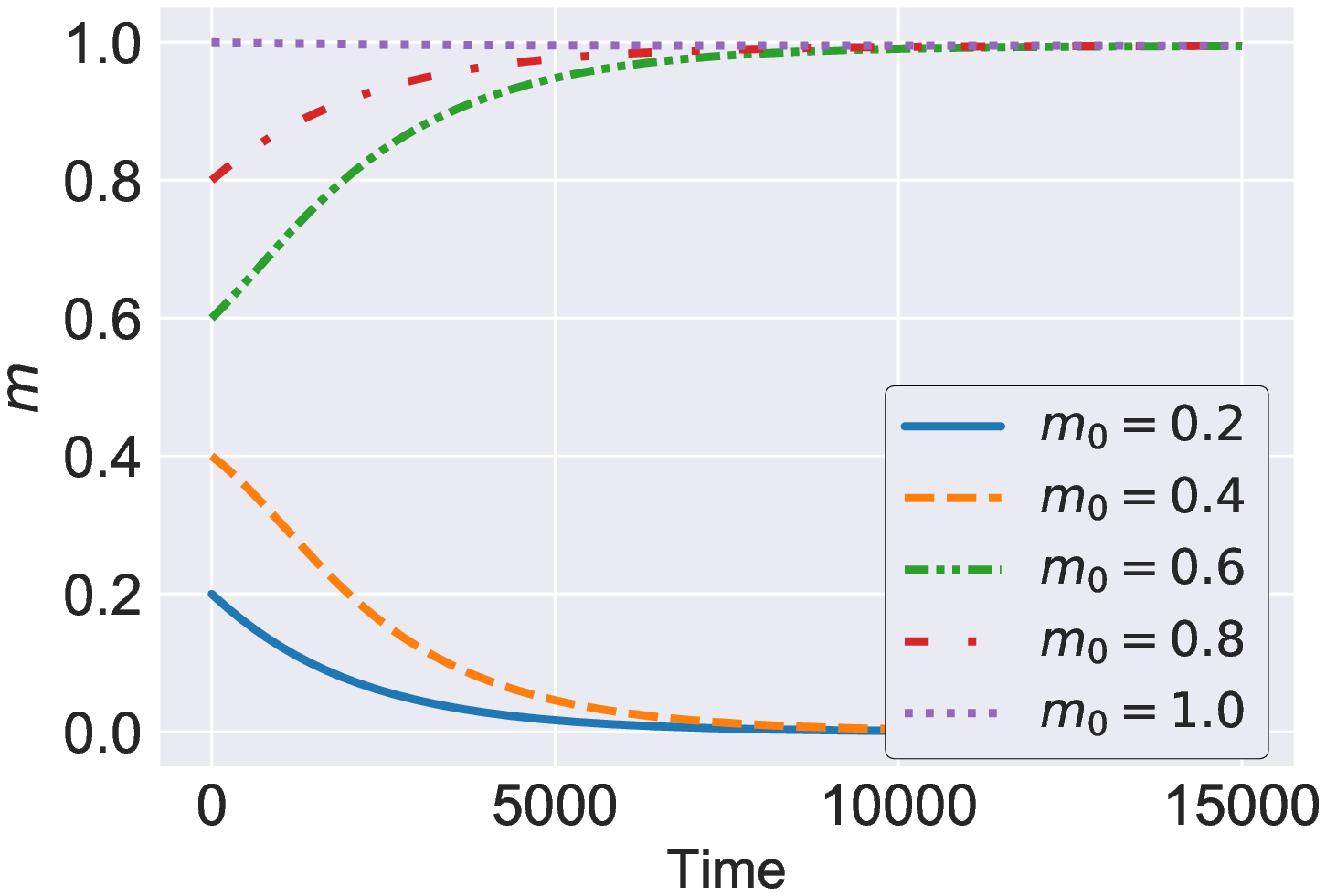}}
\subfigure[\label{fig:fig_1b}]{\includegraphics[width=50mm]{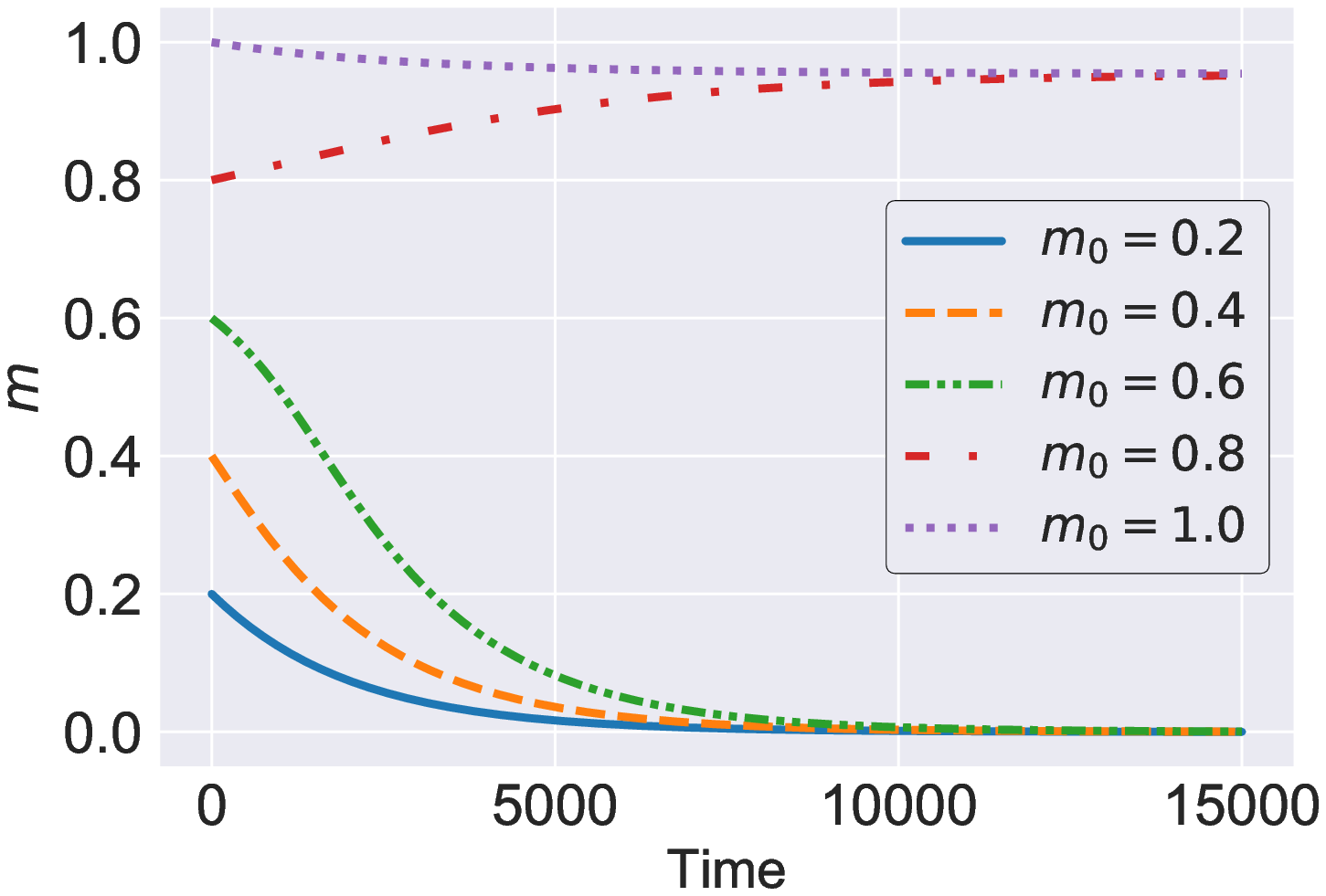}}
\subfigure[\label{fig:fig_1c}]{\includegraphics[width=50mm]{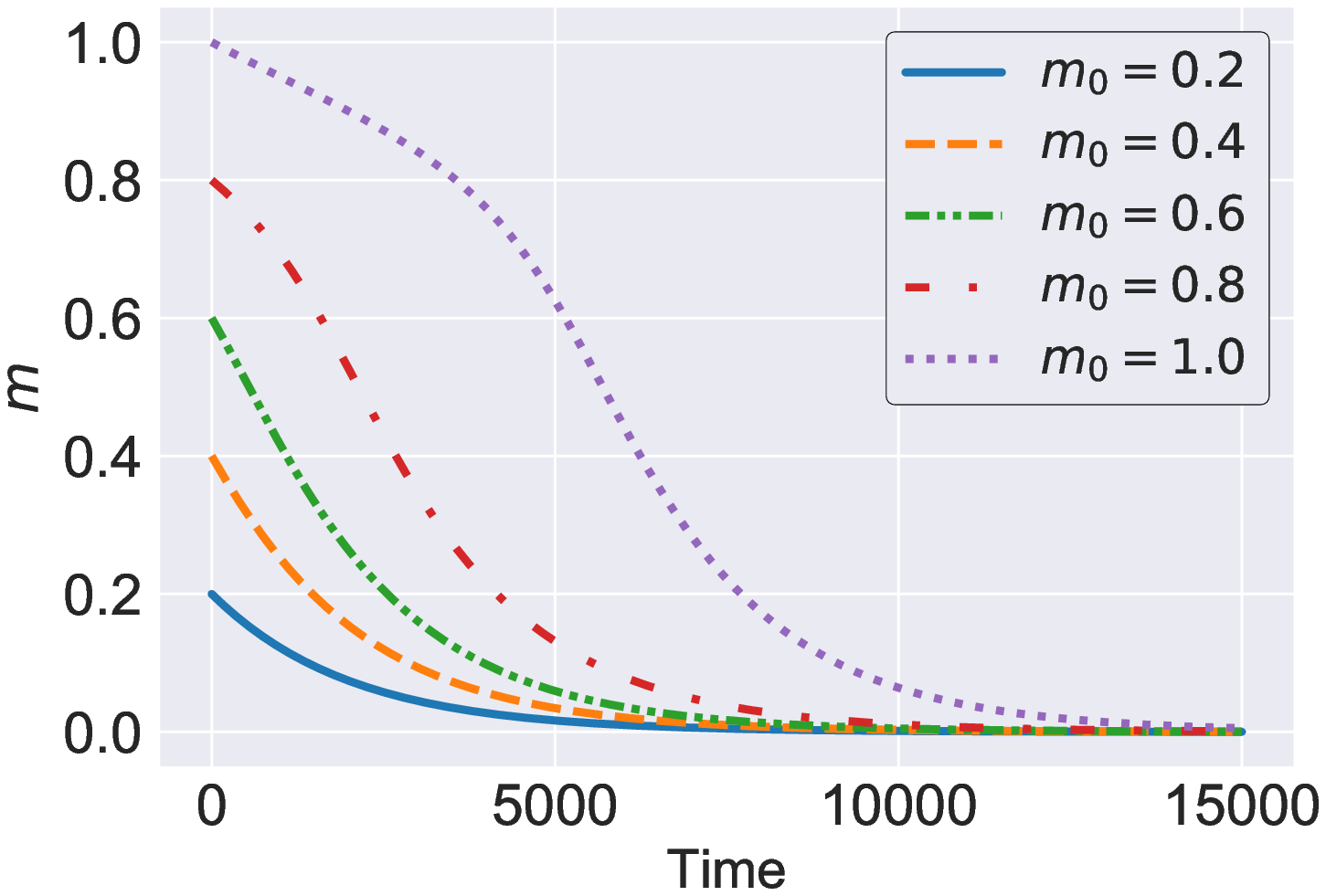}}
\caption{Dynamics of order parameters without anti-ferromagnetic XX interaction  $\gamma=0$, given initial magnetization $m_0=0.2,0.4,0.6,0.8,1.0$ and initial transverse magnetization $m_x=\sqrt{1-m_0^2}$. The horizontal axis denotes the time $t$ of the deterministic flow equation, and the vertical axis denotes the longitudinal magnetization. The experimental settings are (a) $\Gamma=0.5$, (b) $\Gamma=1.3$ and (c) $\Gamma=2.5$. }
\label{fig:fig_1}
\end{figure*}

\begin{figure*}[t]
\subfigure[\label{fig:fig_2a}]{\includegraphics[width=50mm]{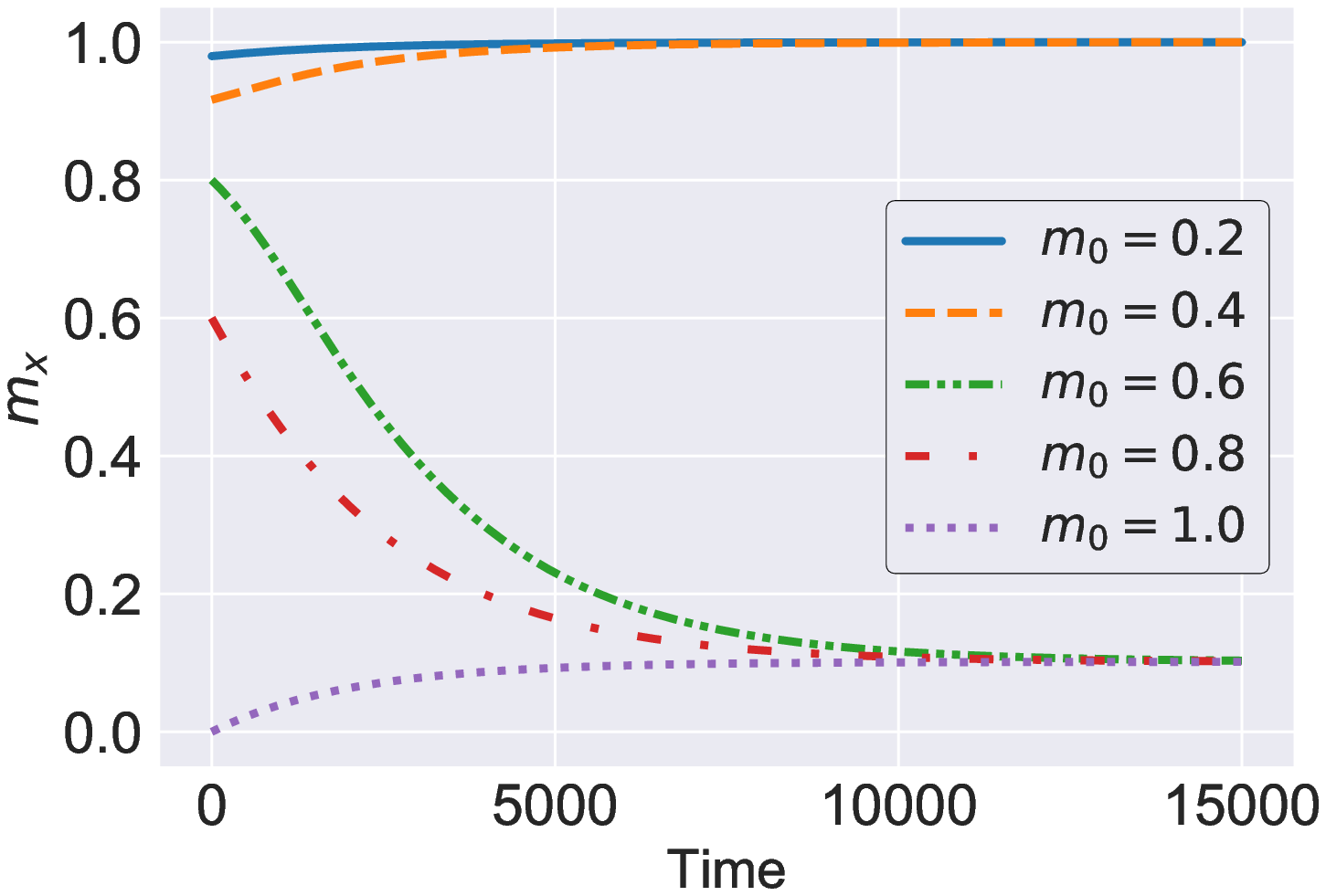}}
\subfigure[\label{fig:fig_2b}]{\includegraphics[width=50mm]{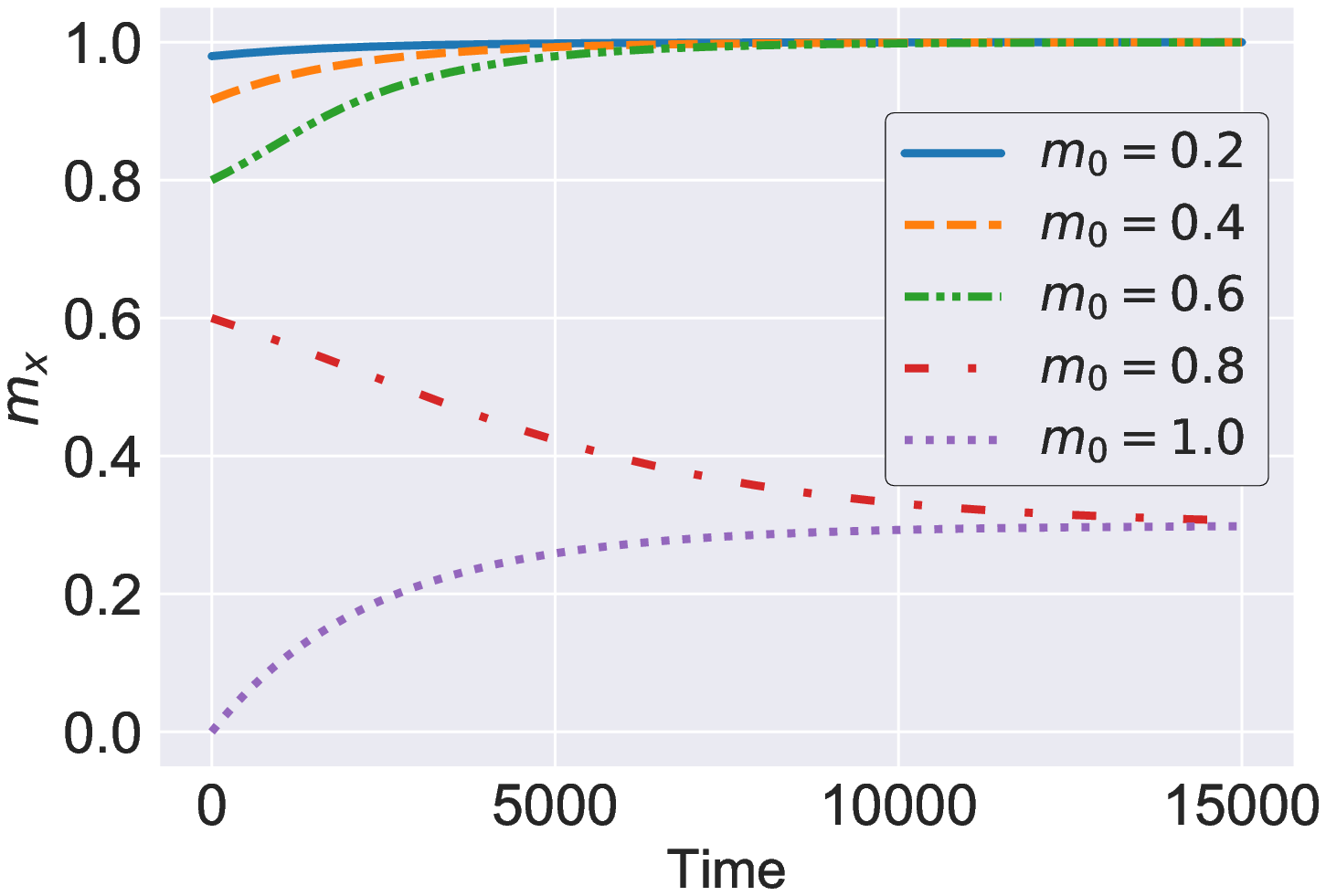}}
\subfigure[\label{fig:fig_2c}]{\includegraphics[width=50mm]{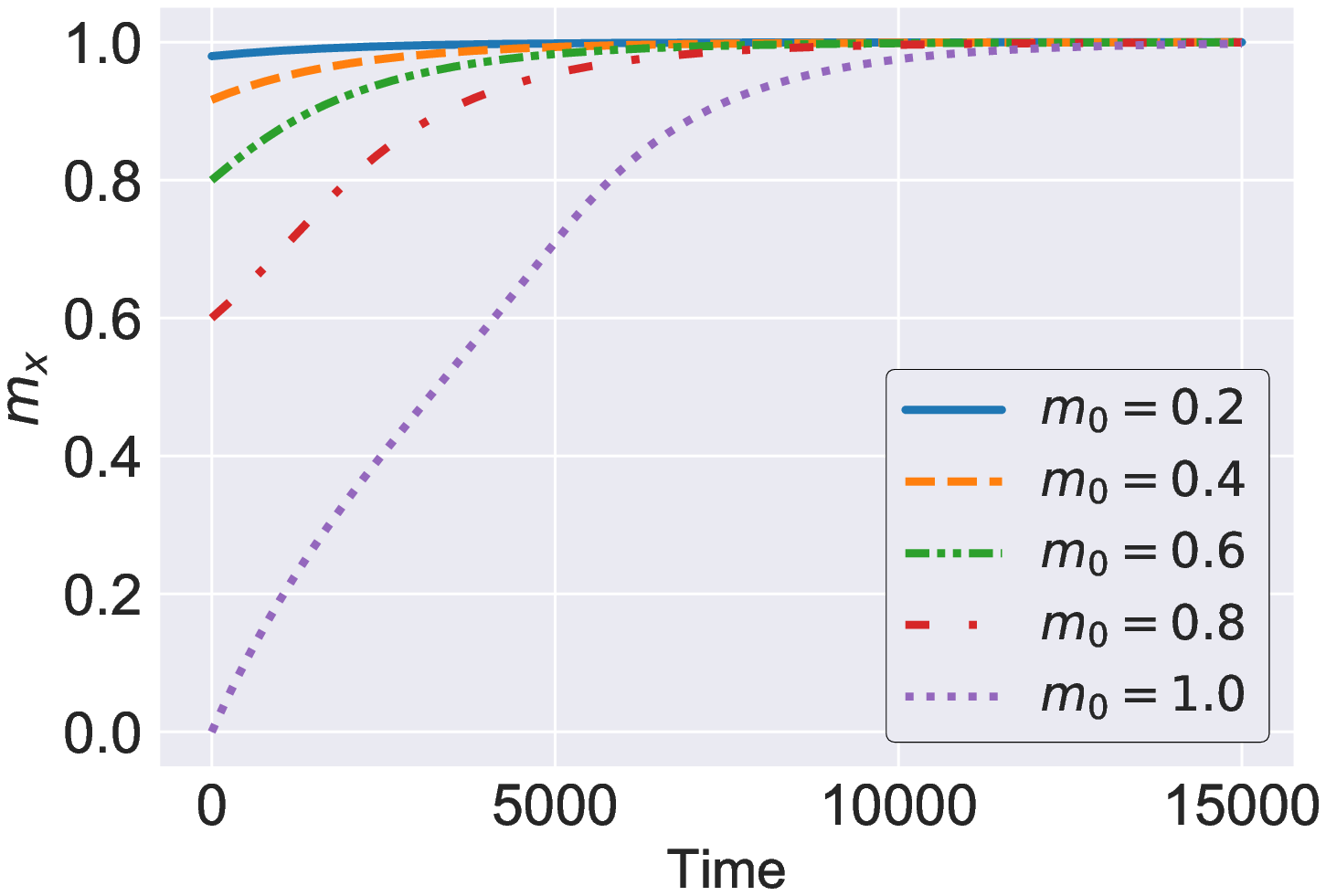}}
\caption{Dynamics of  order parameters without anti-ferromagnetic XX interaction, given initial magnetization $m_0=0.2,0.4,0.6,0.8,1.0$ and initial transverse magnetization $m_x=\sqrt{1-m_0^2}$. The horizontal axis denotes the time $t$ of the deterministic flow equation, and the vertical axis is the transverse magnetization.  The experimental settings are the same as those in Fig. \ref{fig:fig_1}. }
\label{fig:fig_2}
\end{figure*}

\begin{figure*}[t]
\centering
\subfigure[\label{fig:fig_3a}]{\includegraphics[width=50mm]{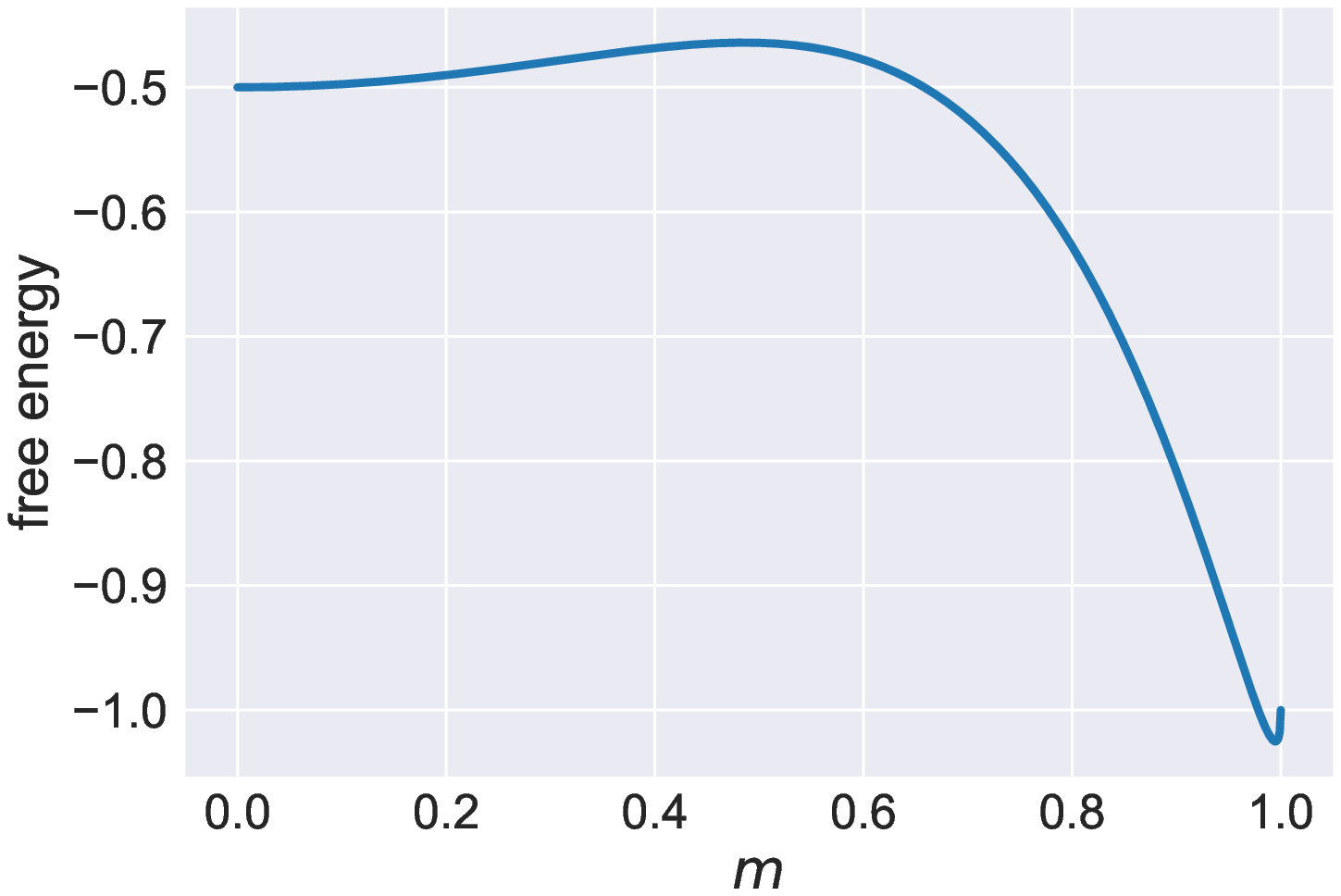}}
\subfigure[	\label{fig:fig_3b}]{\includegraphics[width=50mm]{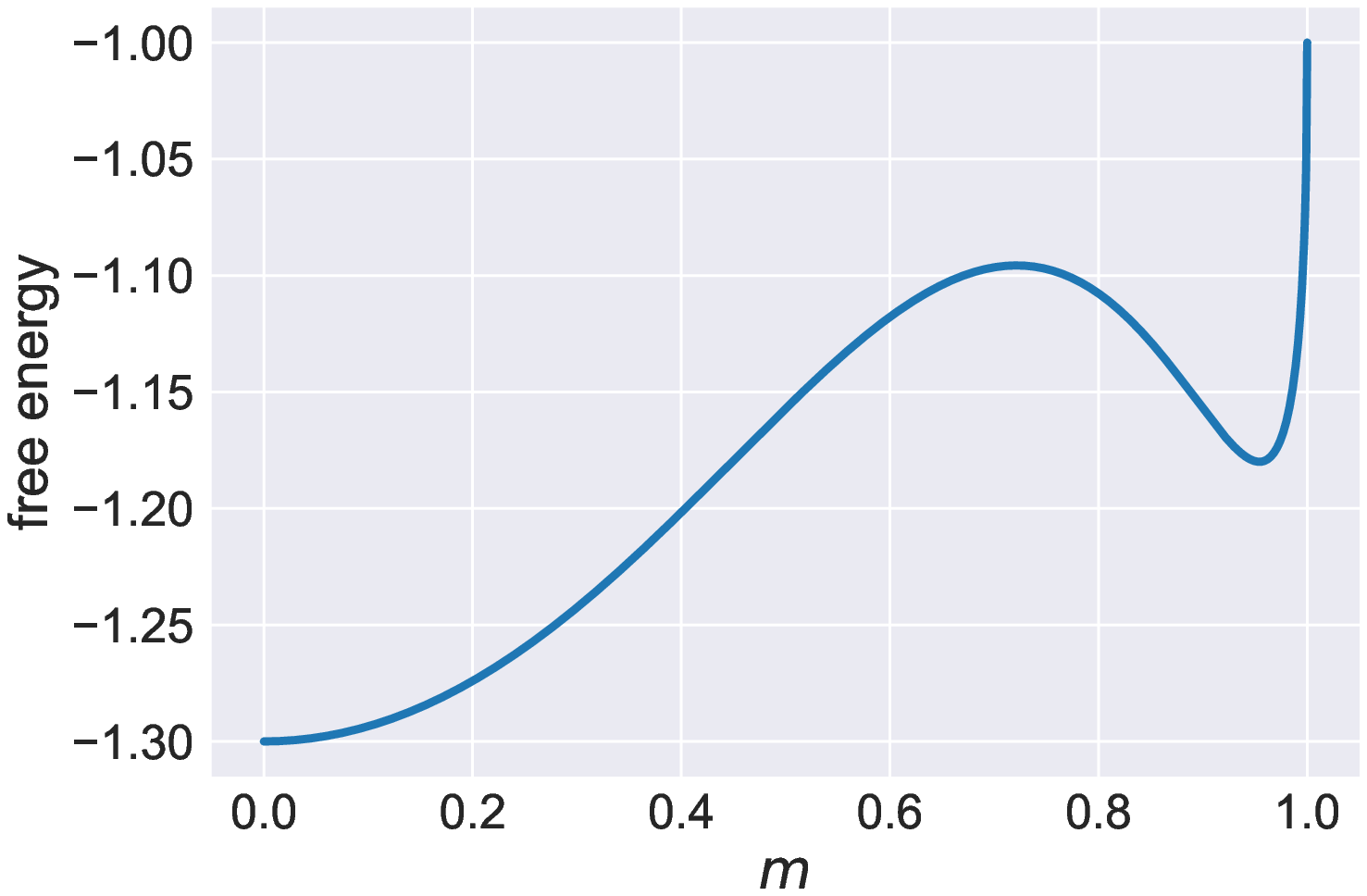}}
\subfigure[\label{fig:fig_3c}]{\includegraphics[width=50mm]{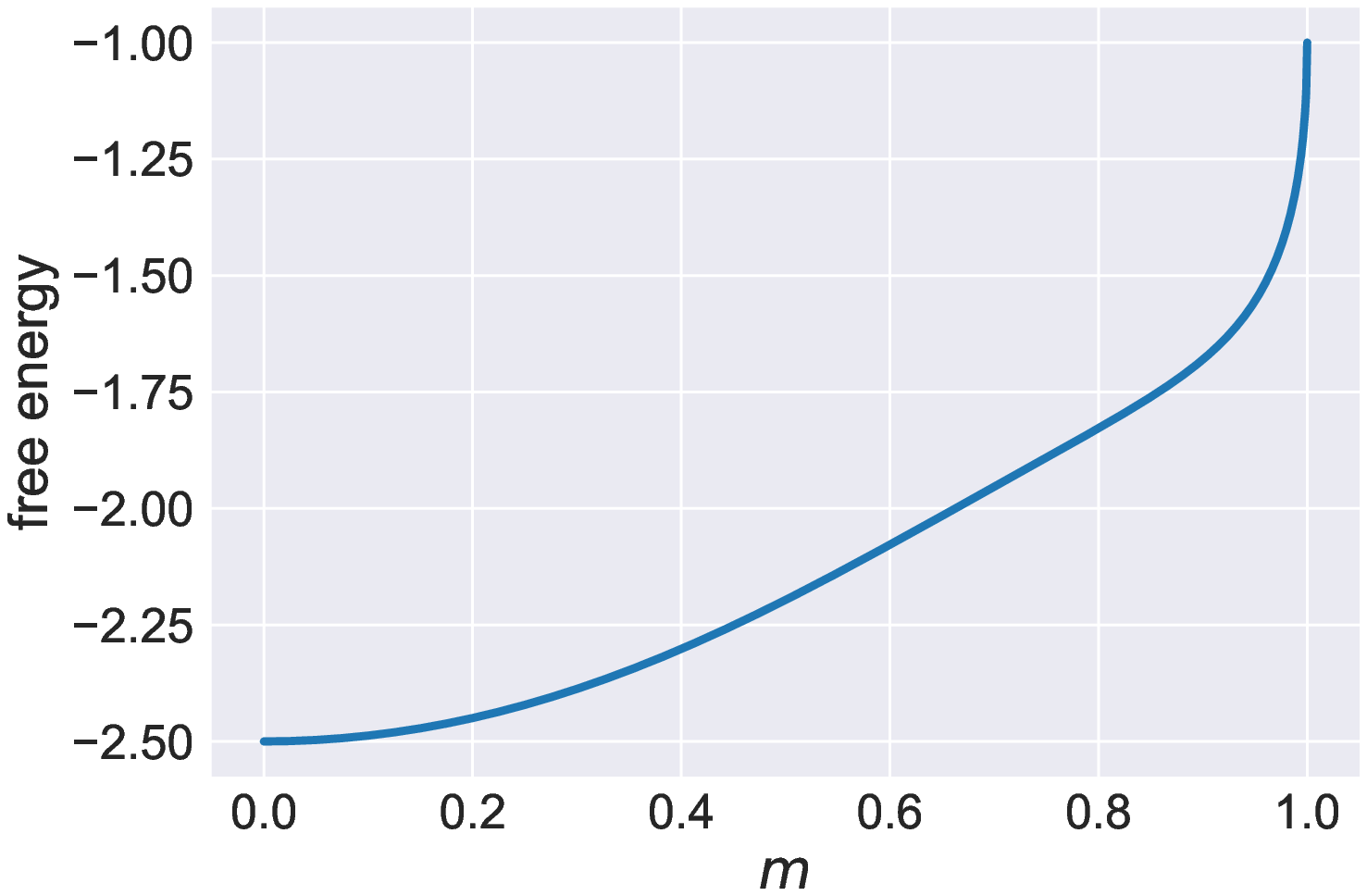}}
\caption{Landscape of the pseudo free energy (\ref{pseudo_f}) without anti-ferromagnetic XX interaction.  The horizontal axis is the longitudinal magnetization, and the vertical axis denotes the pseudo free energy. The experimental settings are the same as those for Figs .\ref{fig:fig_1} and \ref{fig:fig_2}.}
\label{fig:fig_3} 
\end{figure*}
\begin{figure*}[t]
\centering
\subfigure[ \label{fig:fig_4a}Longitudinal magnetization]{\includegraphics[width=50mm]{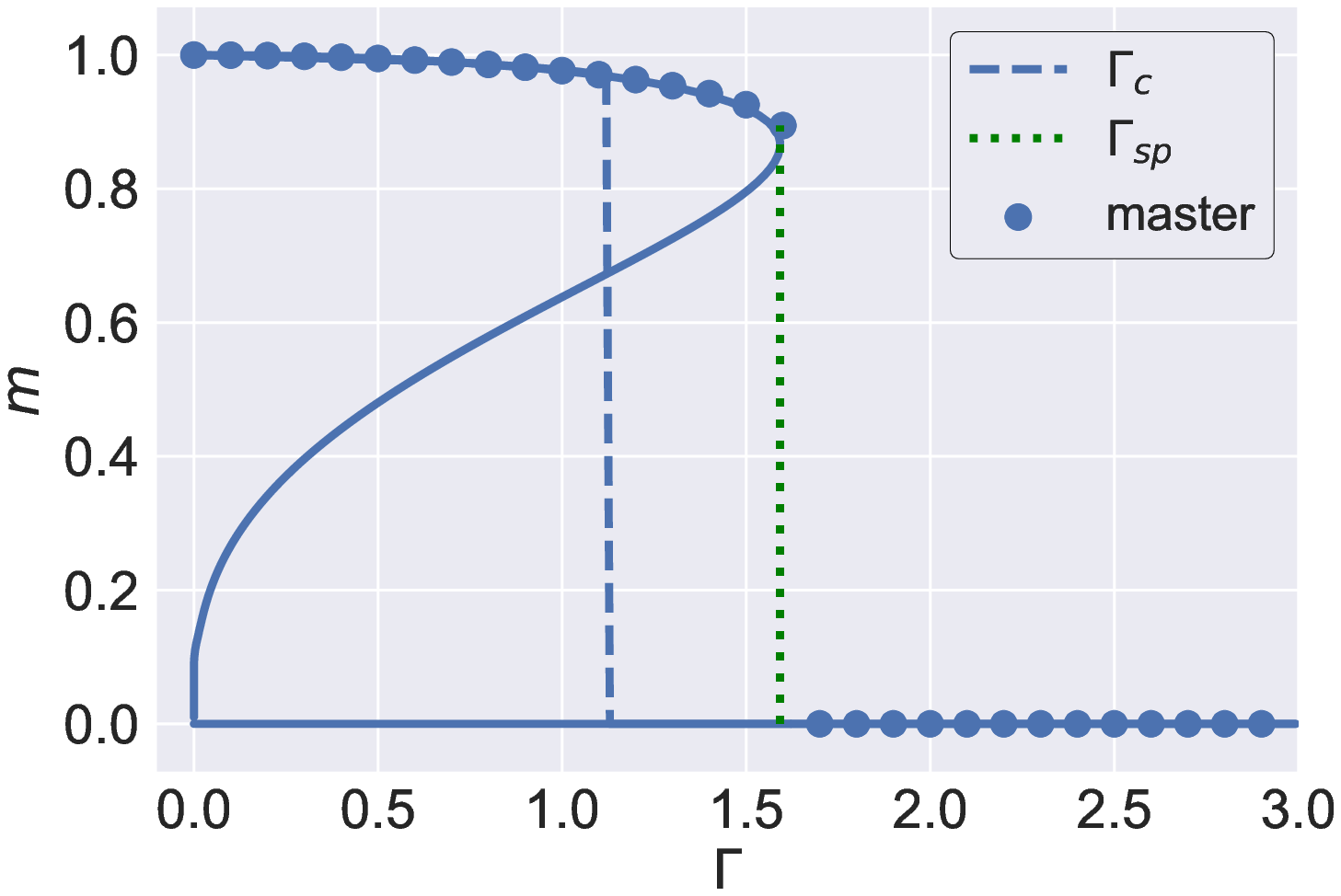}}
\hspace{10mm}
\subfigure[\label{fig:fig_4b}Transverse magnetization]{\includegraphics[width=50mm]{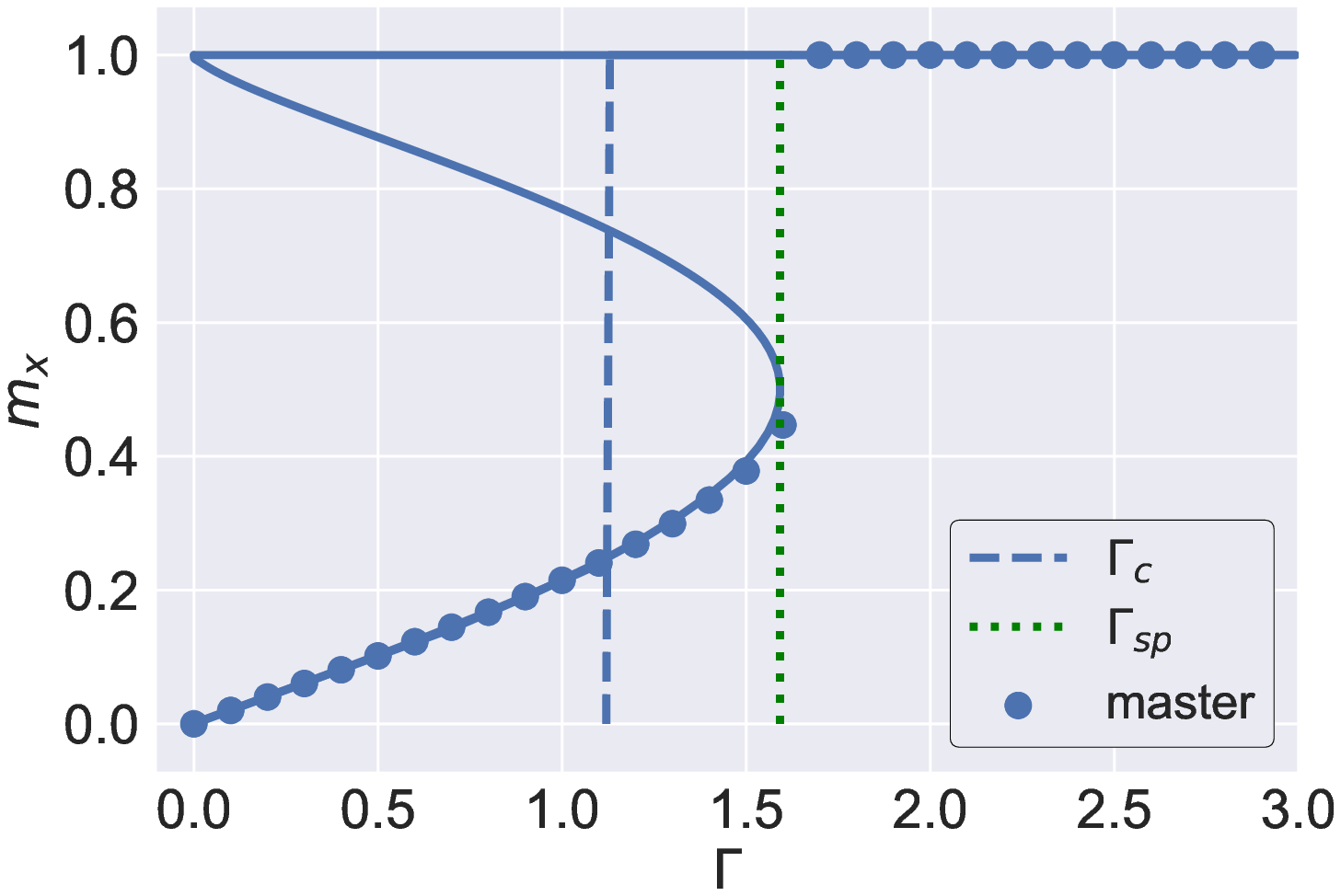}}
\caption{Order parameters without anti-ferromagnetic XX interaction from the master equation and from the saddle-point equations. The figure on the left shows the longitudinal magnetization and that on the right shows the transverse magnetization. The vertical axis denotes these order parameters. The horizontal axis denotes the strength of the transverse magnetic field. The circle denotes the solution obtained by the master equation, whereas the solid line denotes the solution obtained by the saddle-point equations. The blue dashed line denotes the critical point where the pseudo free energy takes the same value. The green dotted line denotes the spinodal point.
}\label{fig:fig_4}
\end{figure*}

 First, we show the dynamics of order parameters without anti-ferromagnetic XX interaction $\gamma=0$ in Figs. \ref{fig:fig_1} and \ref{fig:fig_2}. The original model has a first-order phase transition. We set three different conditions: $\Gamma=0.5$ in the ferromagnetic phase, $\Gamma=2.5$ in the paramagnetic phase, and $\Gamma=1.3$ in the paramagnetic phase between the critical point $\Gamma_c$ and the spinodal point $\Gamma_{sp}$. These figures indicate that the dynamics exponentially converge to each steady state  depending on the initial condition.
 
 We also consider the pseudo free energy to evaluate the equilibrium solutions. By using mean-field theory, 
the pseudo free energy is written as 
\begin{align}
&F(m,\tilde{m},m_x,\tilde{m}_x,\beta,\Gamma,\gamma)=-m^p+m\tilde{m}\nonumber \\
&-\Gamma \tilde{m}_x+\frac{\gamma}{2}m_x^2+m_x\tilde{m}_x-\frac{1}{\beta}\log2\cosh \left(\beta \sqrt{(\tilde{m})^2+(\tilde{m}_x)^2}\right). 
\label{pseudo_f}
\end{align}
 Standardly, with mean-field theory, we utilize the saddle-point conditions $\partial F/\partial m=0$ and $\partial F/\partial m_x=0$ for $\tilde{m}$ and $\tilde{m}_x$, respectively. 
 Strictly speaking, however, these conditions need not be used to estimate the pseudo free energy precisely (\ref{pseudo_f}). Therefore, we utilize the saddle-point conditions  $\partial F/\partial \tilde{m}=0$ and $\partial F/\partial \tilde{m}_x=0$. These conditions lead to 
  \begin{align}
m=\frac{\tilde{m}}{\sqrt{\tilde{m}^2+\tilde{m}_x^2}}\tanh \left(\beta \sqrt{\tilde{m}^2+\tilde{m}_x^2}\right),\\
m_x=\frac{\tilde{m}_x}{\sqrt{\tilde{m}^2+\tilde{m}_x^2}}\tanh \left(\beta \sqrt{\tilde{m}^2+\tilde{m}_x^2}\right).
\end{align}
Even if we utilize these saddle-point conditions, we can ultimately obtain the saddle-point equations as
 \begin{align}
m&=\left(pm^{p-1}\right)\frac{\tanh \left(\beta \sqrt{\left(pm^{p-1}\right)^2+\left(\Gamma-\gamma m_x\right)^2}\right)}{ \sqrt{\left(pm^{p-1}\right)^2+\left(\Gamma-\gamma m_x\right)^2} 
},\\
m_x&=\left(\Gamma-\gamma m_x\right)\frac{\tanh \left(\beta \sqrt{\left(pm^{p-1}\right)^2+\left(\Gamma-\gamma m_x\right)^2}\right)}{ \sqrt{\left(pm^{p-1}\right)^2+\left(\Gamma-\gamma m_x\right)^2} 
}.
\end{align}
These saddle-point equations, obtained in the standard manner for mean-field theory \cite{seki_antiferro}, 
 are consistent with $dm/dt=0$ and $dm_x/dt=0$.
 
 To show the validity of the solution obtained from the master equation, we plot the pseudo free energy (\ref{pseudo_f}) with respect to the function of longitudinal magnetization $m$ in Fig.\ref{fig:fig_3}. Here, we can utilize the equation $m^2+m_x^2=1$ in the thermodynamic limit $N \rightarrow \infty$. According to the initial values, each equilibrium solution from the master equation converges to the minimum values. 
From Fig.\ref{fig:fig_3a}, the pseudo free energy has two different stable values $m\simeq1$ and $m	\simeq0$. The solution $m\simeq0$ is the metastable state in the ferromagnetic phase. 
Therefore, we can see that the equilibrium solutions in Fig.\ref{fig:fig_1a} converge to two different values, $m\simeq1$ and $m\simeq0$, according to the different initial values.  In the paramagnetic phase between the critical point and the spinodal point, a similar phenomenon obtains.
\begin{figure*}[t]
\centering
\subfigure[\label{fig:fig_5a} ]{\includegraphics[width=50mm]{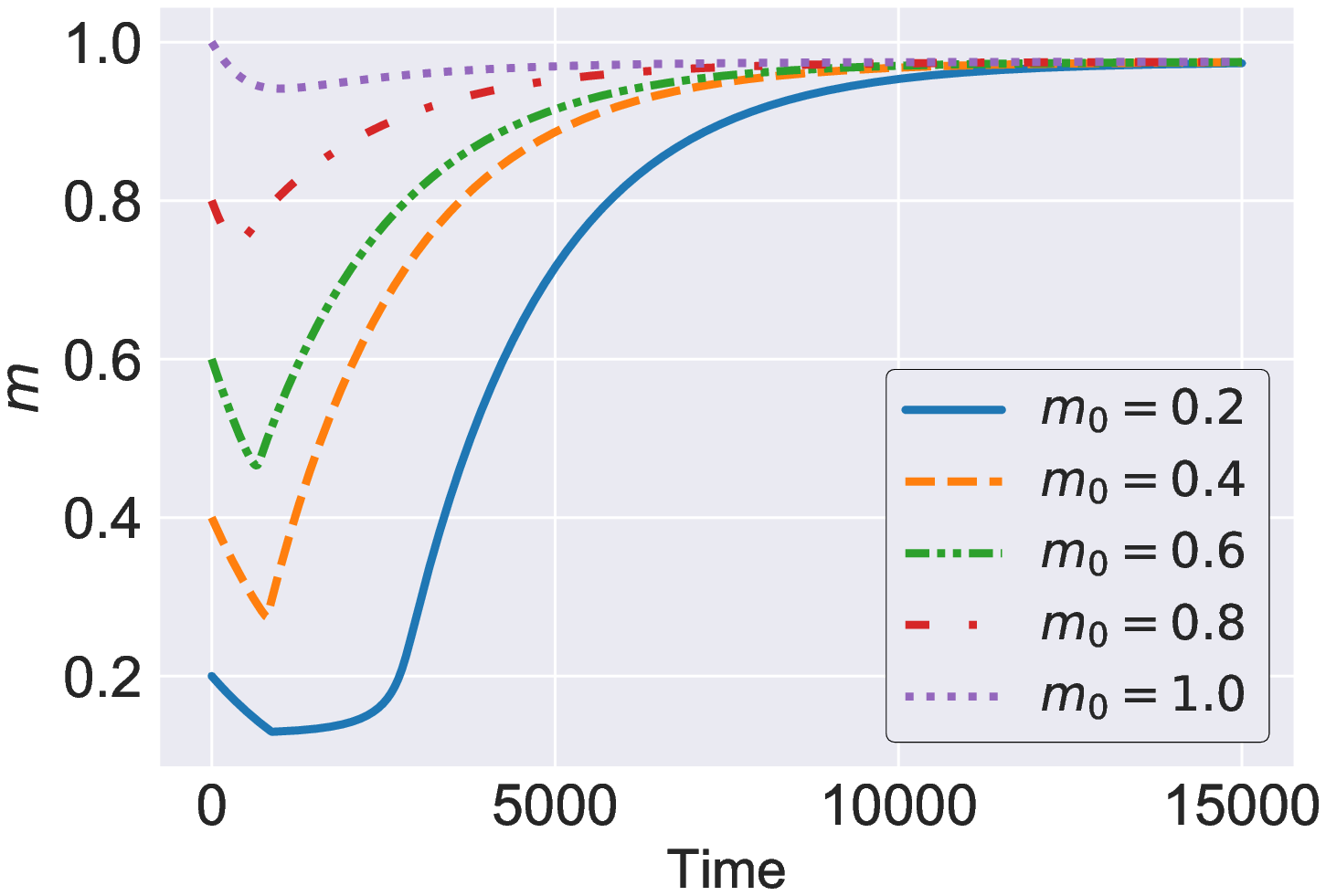}}
\subfigure[\label{fig:fig_5b}]{\includegraphics[width=50mm]{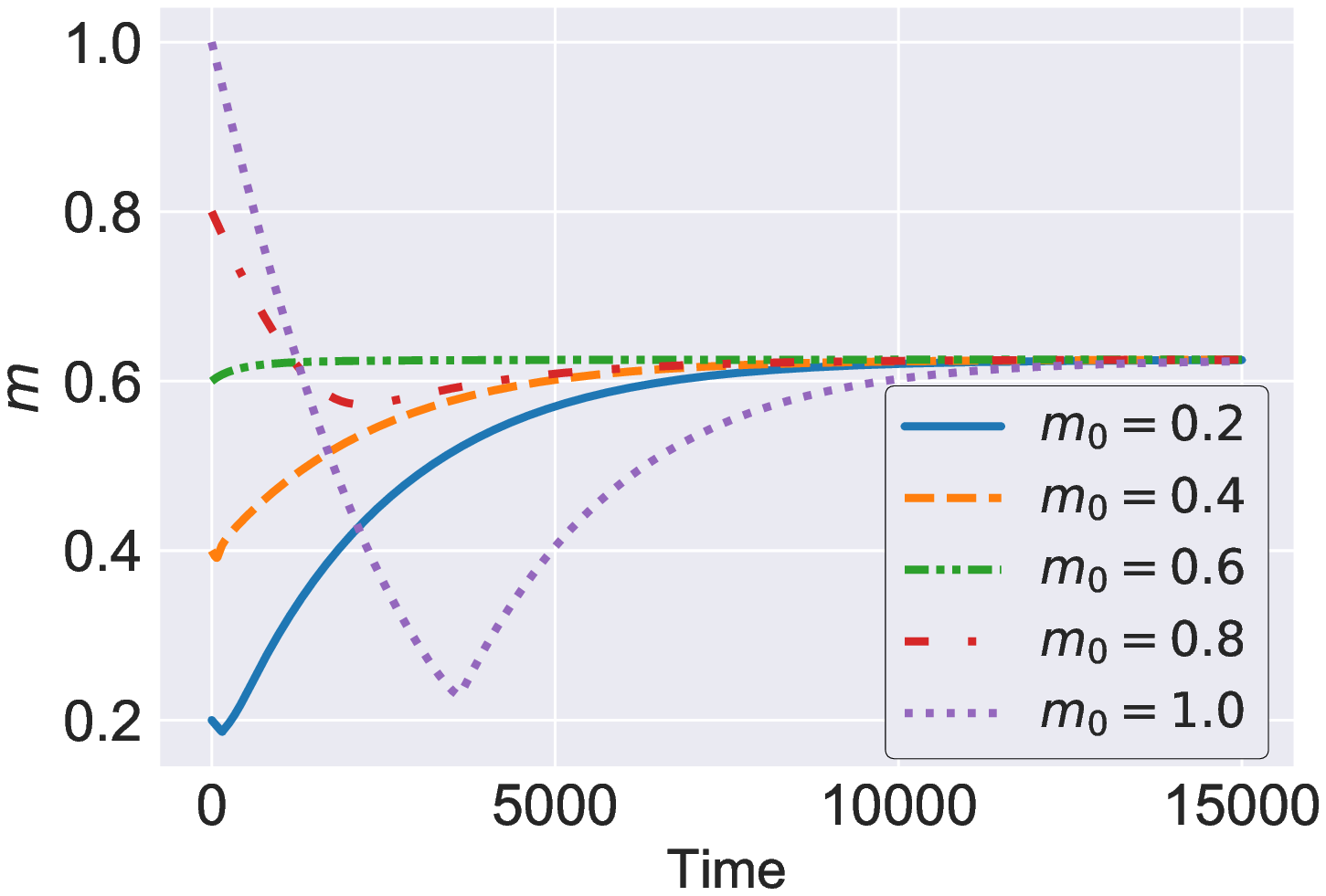}}
\subfigure[\label{fig:fig_5c}]{\includegraphics[width=50mm]{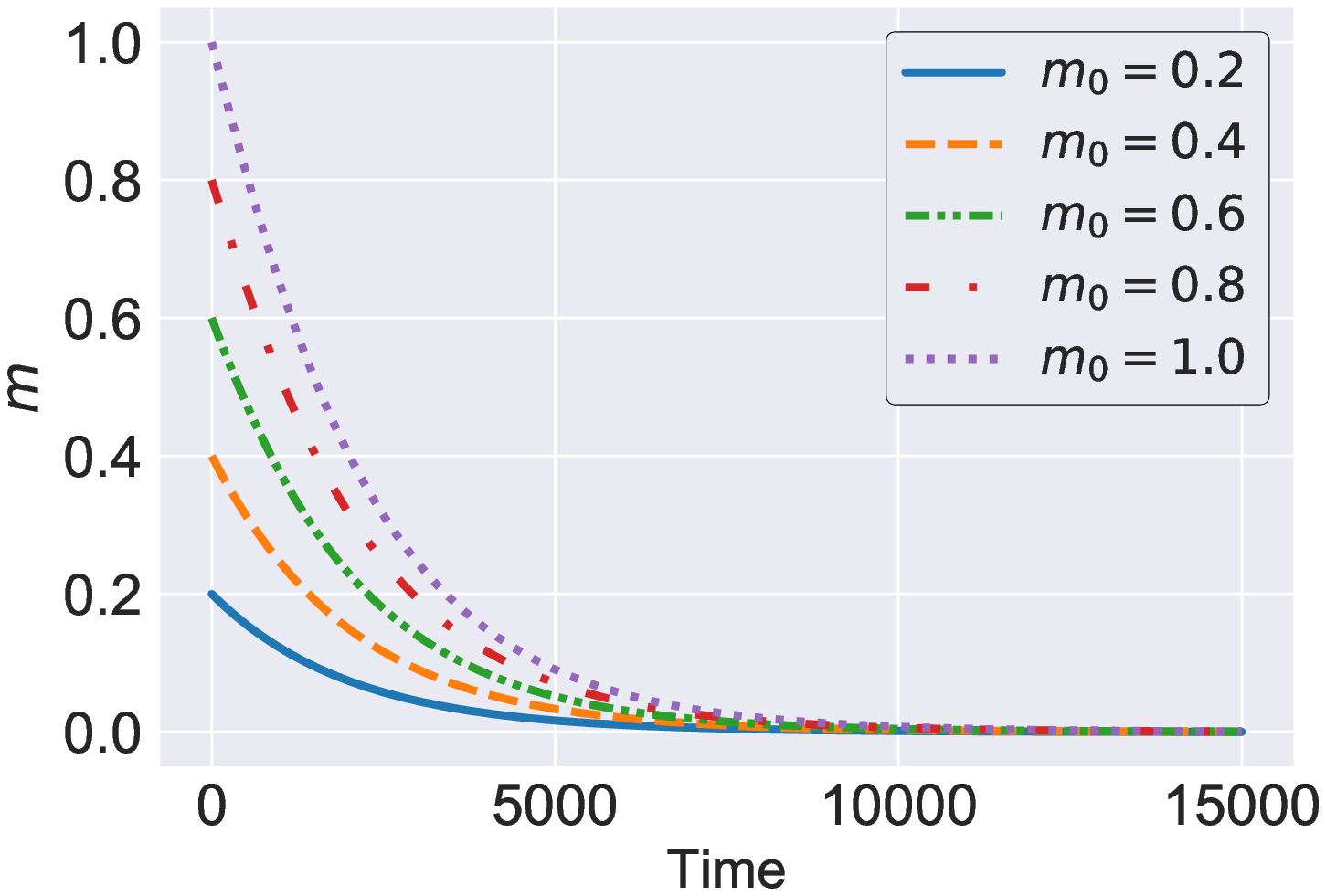}}
\caption{Dynamics of  order parameters with  anti-ferromagnetic XX interaction $\gamma=18$, given initial magnetization $m_0=0.2,0.4,0.6,0.8,1.0$ and initial transverse magnetization $m_x=\sqrt{1-m_0^2}$. The horizontal axis denotes the time $t$ of the deterministic flow equation, and the vertical axis denotes the longitudinal magnetization. The experimental settings are (a) $\Gamma=5$, (b) $\Gamma=15$ and (c) $\Gamma=25$. }
\label{fig:fig_5} 
\end{figure*}

\begin{figure*}[t]
\centering
\subfigure[\label{fig:fig_6a}]{\includegraphics[width=50mm]{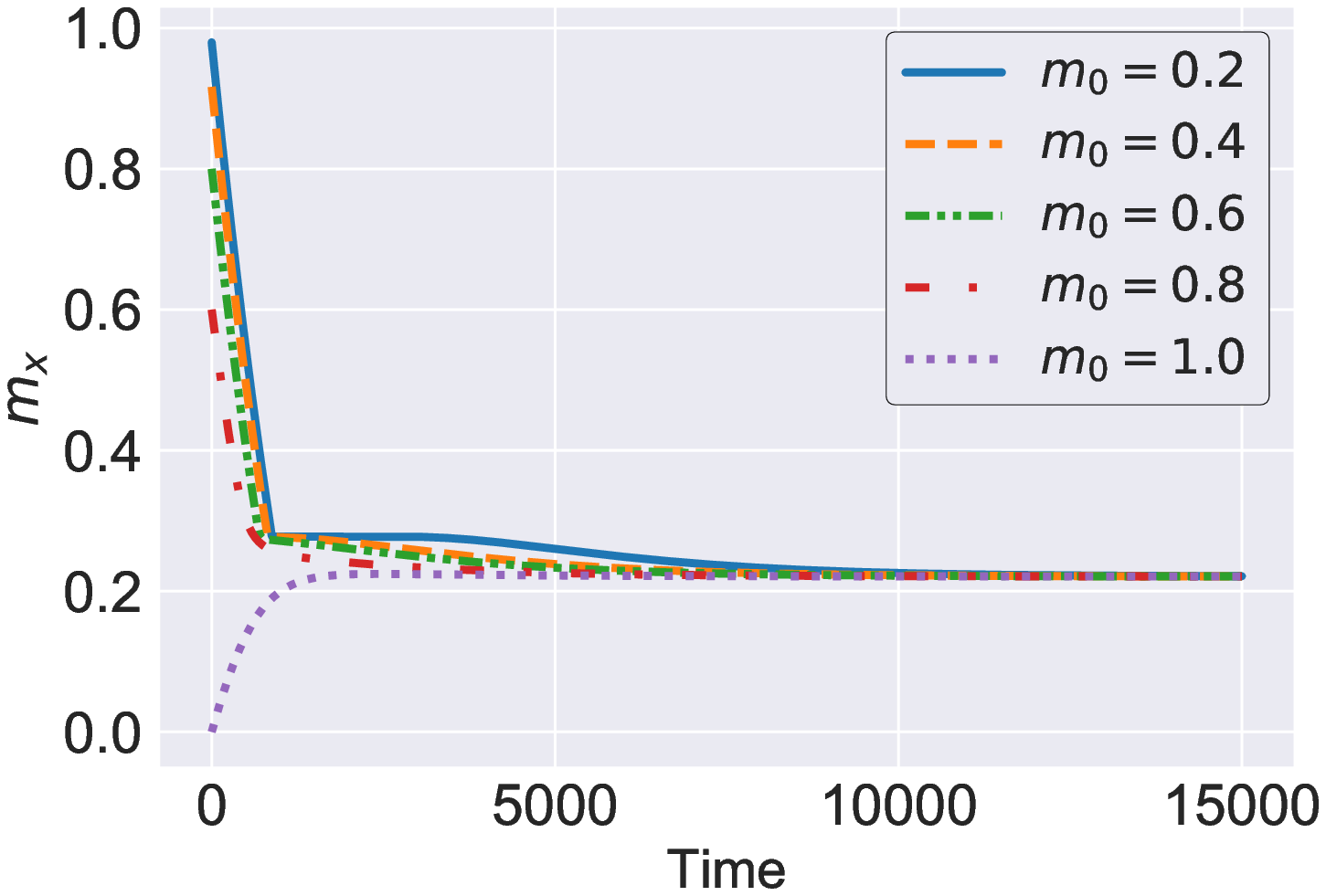}}
\subfigure[\label{fig:fig_6b}]{\includegraphics[width=50mm]{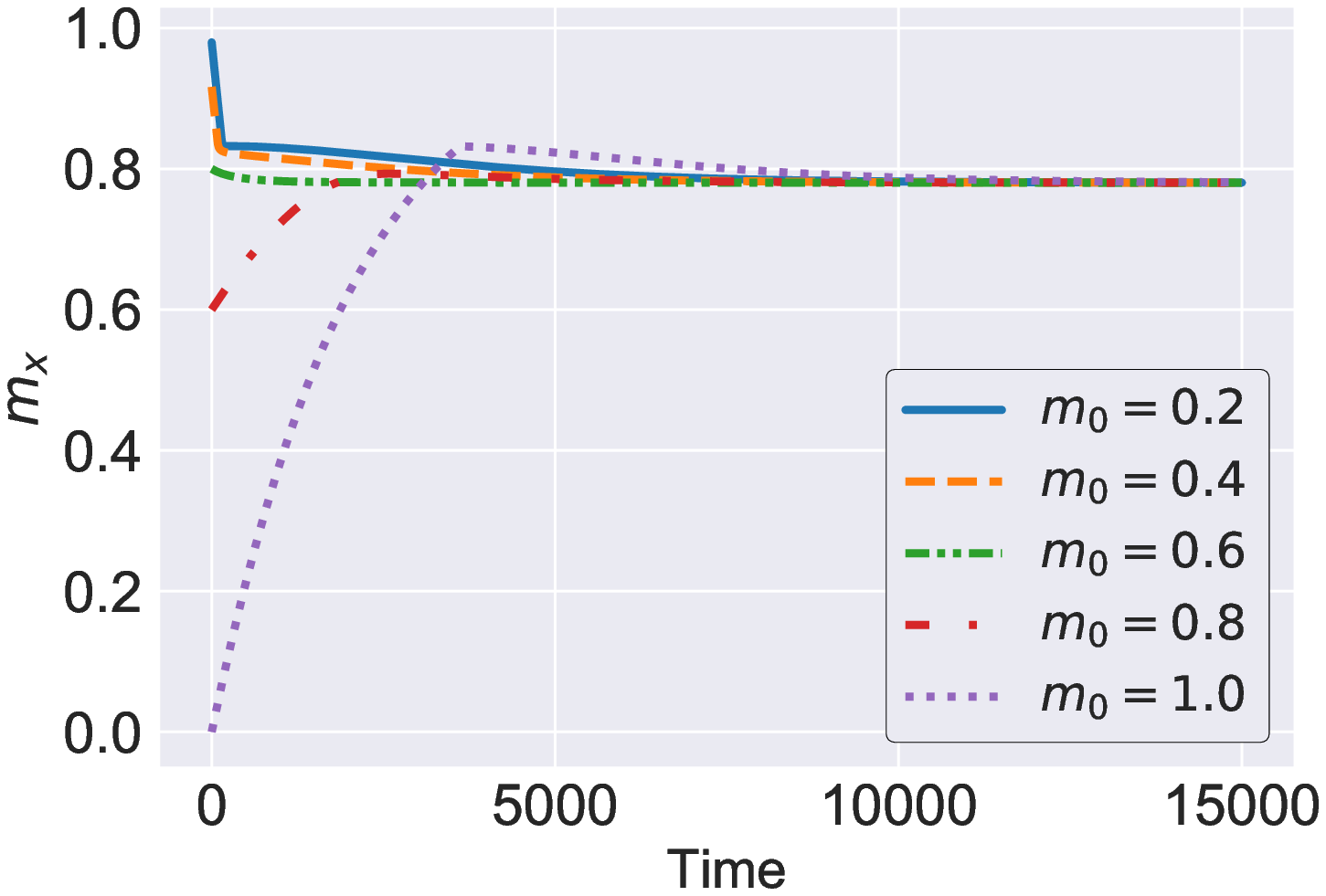}}
\subfigure[\label{fig:fig_6c}]{\includegraphics[width=50mm]{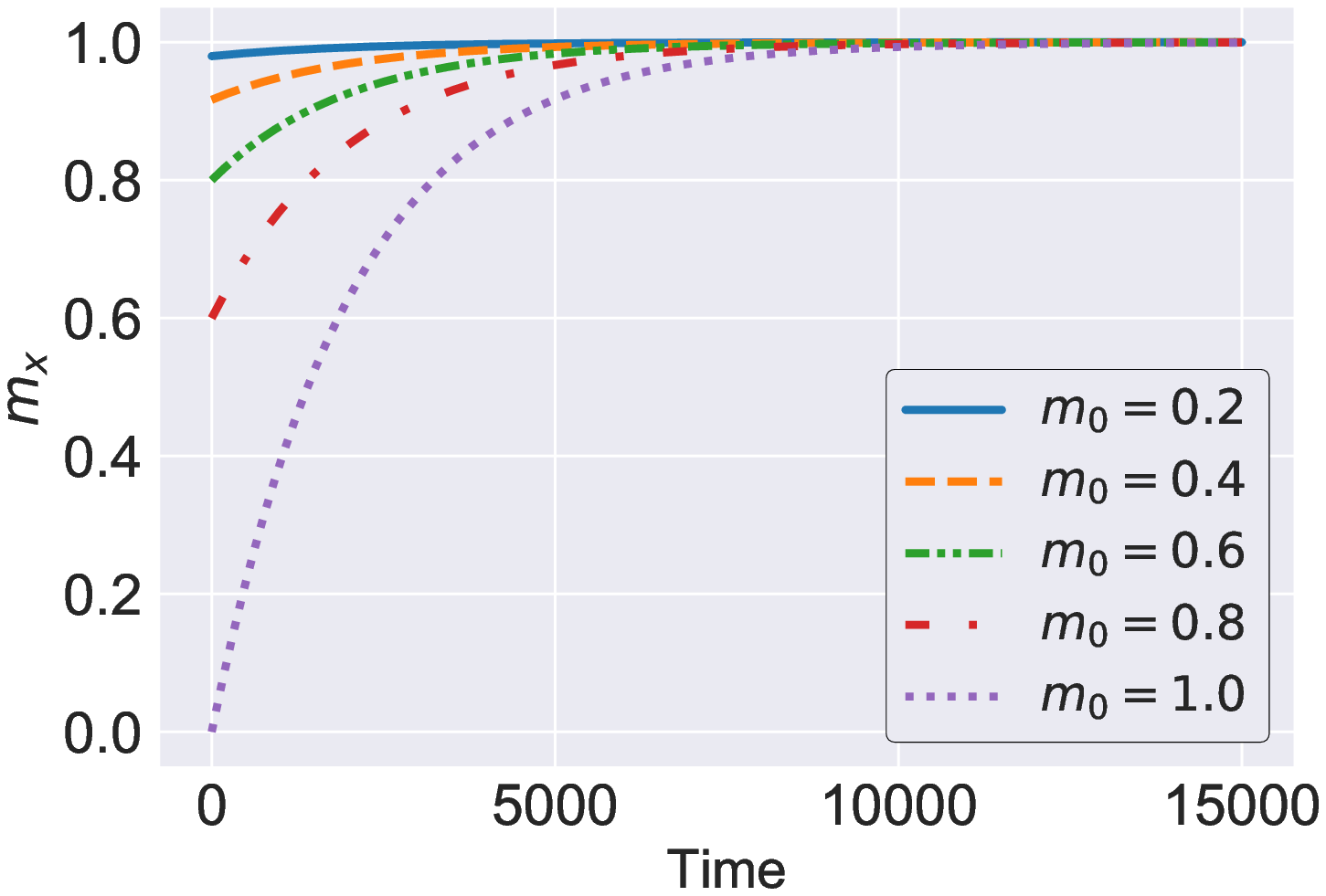}}
\caption{Dynamics of  order parameters with  anti-ferromagnetic XX interaction, given initial magnetization $m_0=0.2,0.4,0.6,0.8,1.0$ and initial transverse magnetization $m_x=\sqrt{1-m_0^2}$. The horizontal axis denotes the time $t$ of the deterministic flow equation, and the vertical axis is the transverse magnetization. The experimental settings are the same as those in Fig. \ref{fig:fig_5}.}
\label{fig:fig_6} 
\end{figure*}

We plot the equilibrium solutions from the master equation and the exact solutions from the saddle-point equations in Fig. \ref{fig:fig_4}. We find that these order parameters change discontinuously. The longitudinal magnetization is the multivalued function with respect to the strength of the transverse magnetic field. After the spinodal point, the ferromagnetic stable state $m>0$ appears. The dashed line in Fig. \ref{fig:fig_4} denotes the critical point where the pseudo free energy takes the same value. 
From the viewpoint of dynamics, we can confirm that this model has a first-order phase transition. 
\begin{figure*}[t]
\centering
\subfigure[\label{fig:fig_7a}]{\includegraphics[width=50mm]{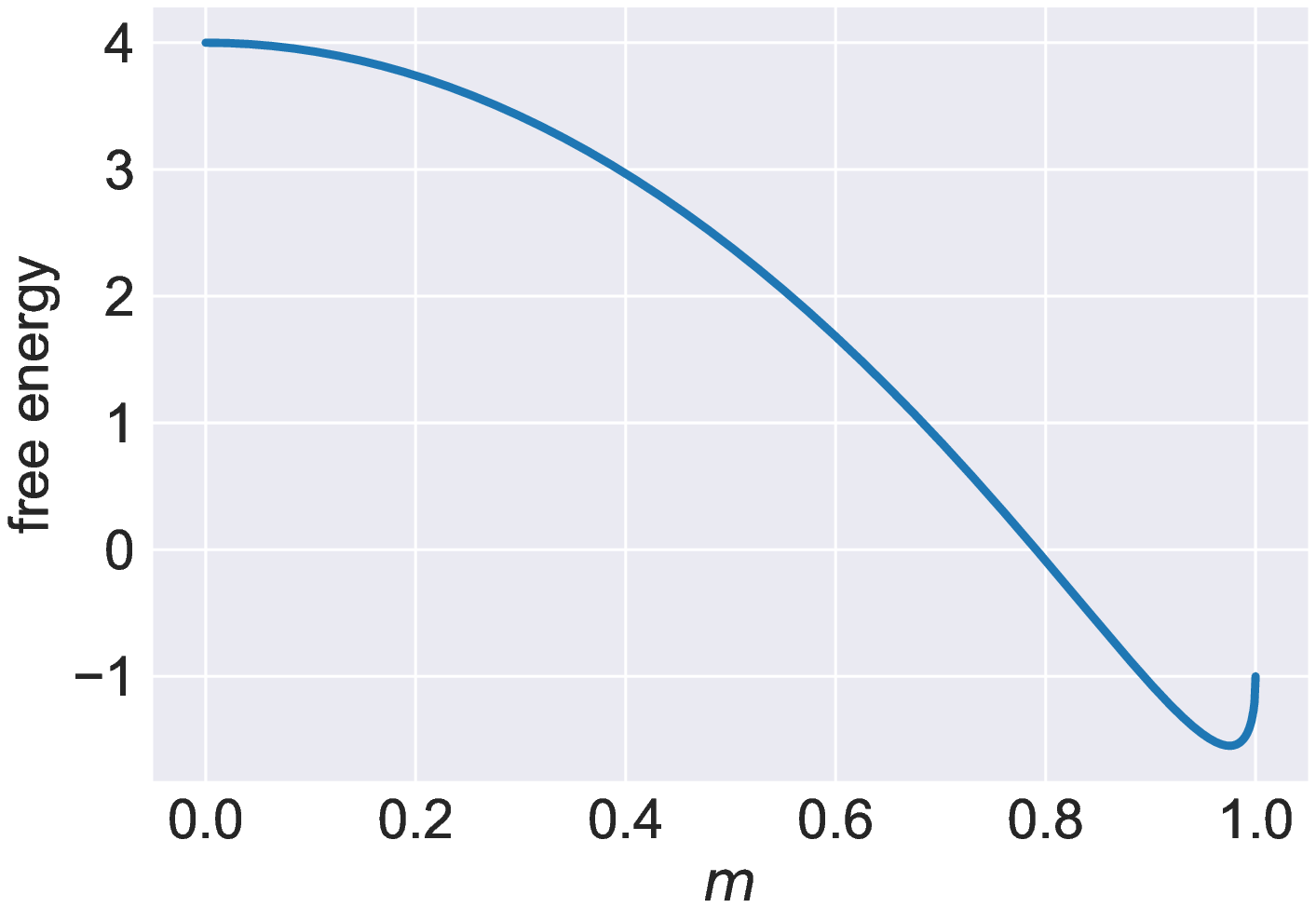}}
\subfigure[\label{fig:fig_7b}]{\includegraphics[width=50mm]{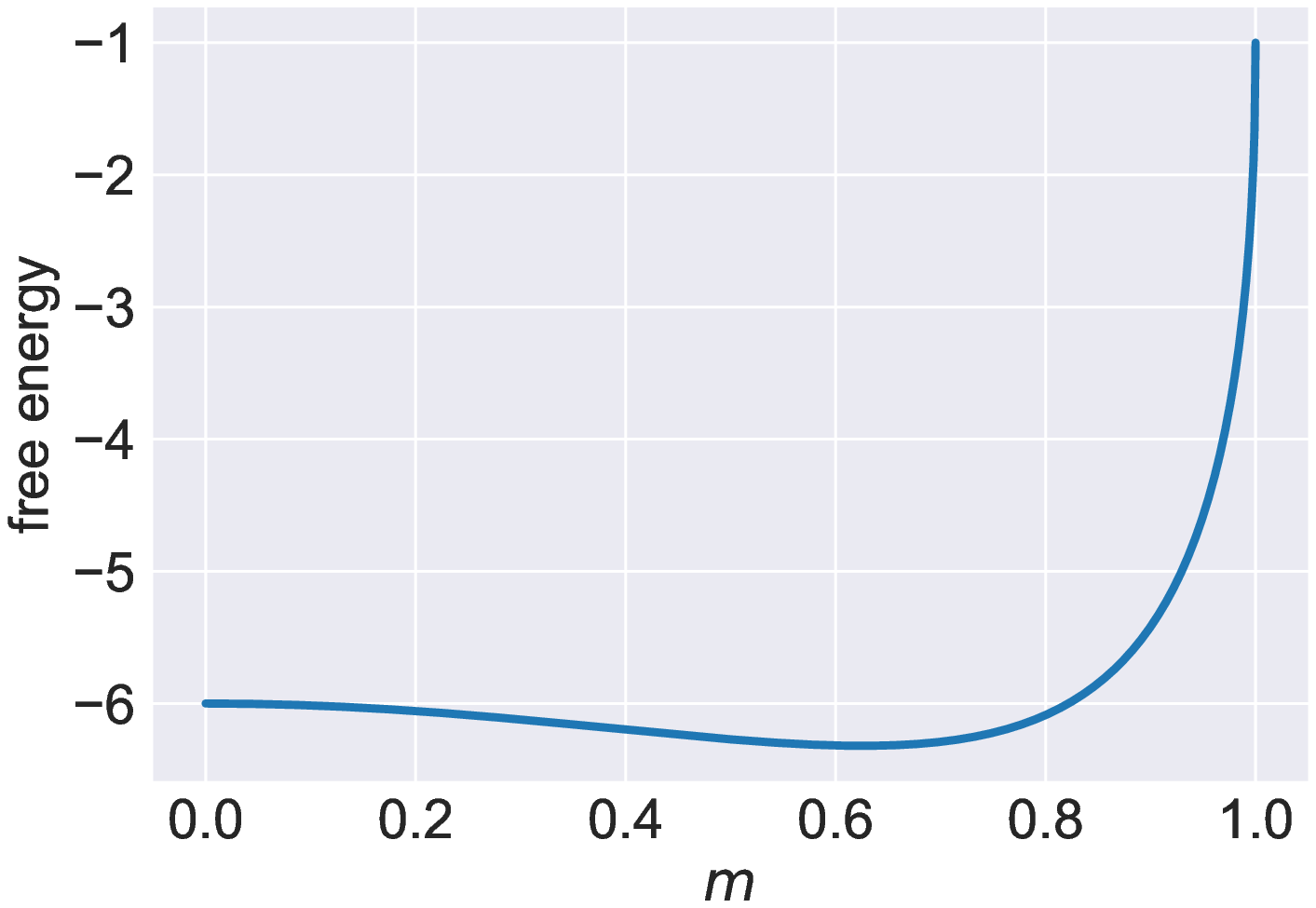}}
\subfigure[\label{fig:fig_7c}]{\includegraphics[width=50mm]{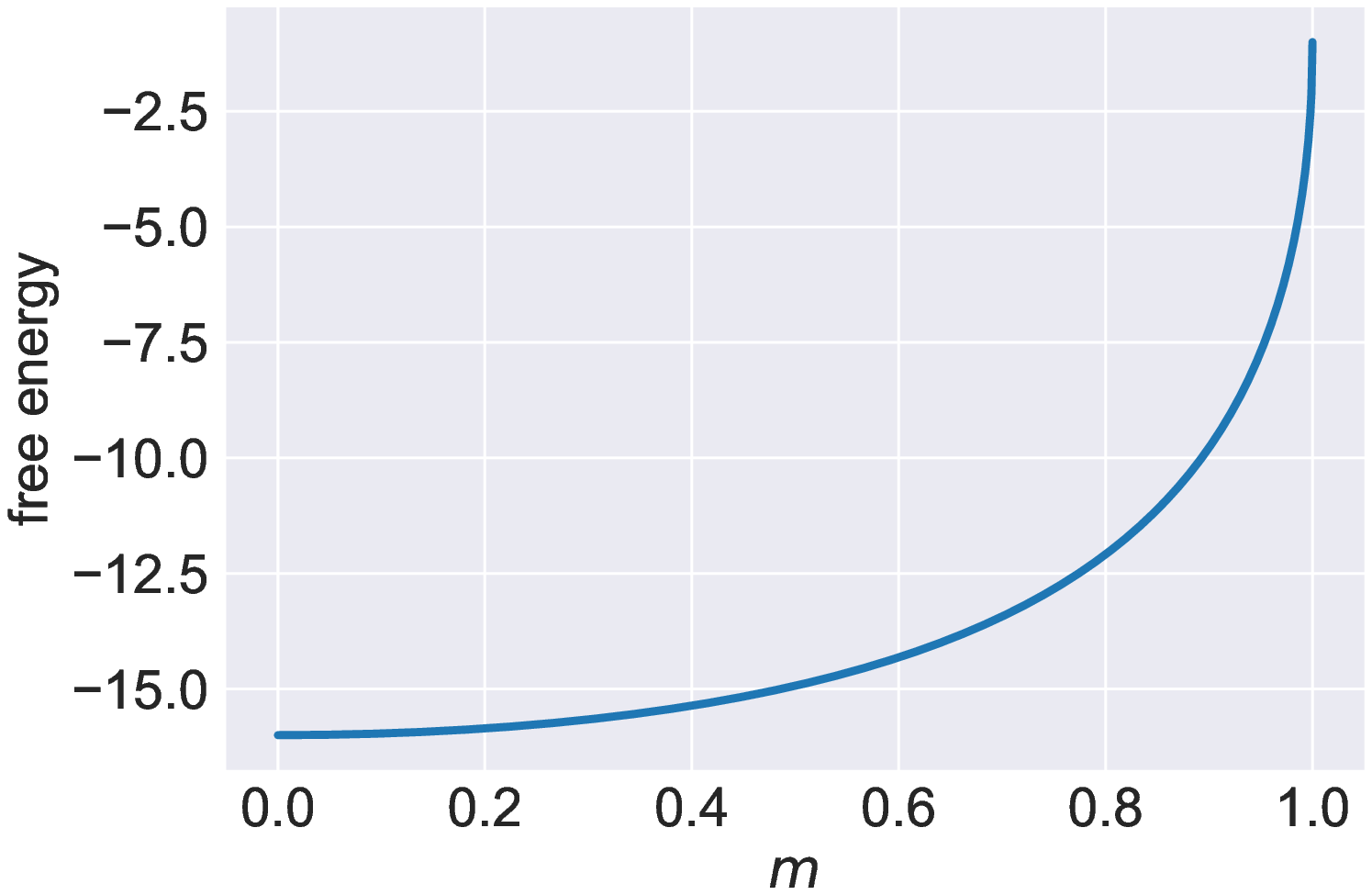}}
\caption{Landscape of the pseudo free energy (\ref{pseudo_f}) with anti-ferromagnetic XX interaction.  The horizontal axis is the longitudinal magnetization. The vertical axis is the pseudo free energy. The experimental settings are the same as those in Figs. \ref{fig:fig_5} and \ref{fig:fig_6}.}
\label{fig:fig_7} 
\end{figure*}

\begin{figure*}[t]
\centering
\subfigure[ \label{fig:fig_8a} Longitudinal magnetization]{\includegraphics[width=50mm]{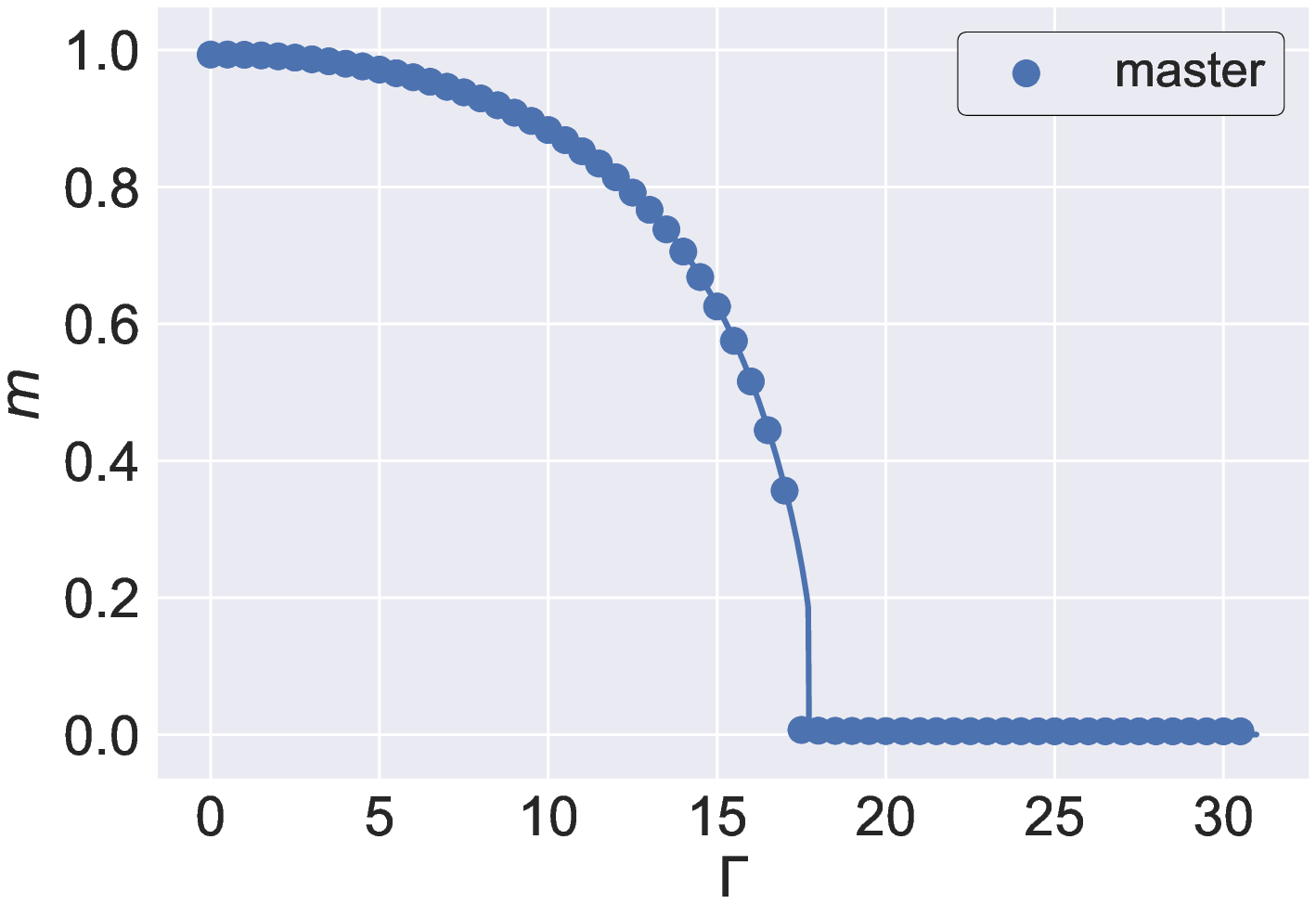}}
\hspace{10mm}
\subfigure[ \label{fig:fig_8b} Transverse magnetization]{\includegraphics[width=50mm]{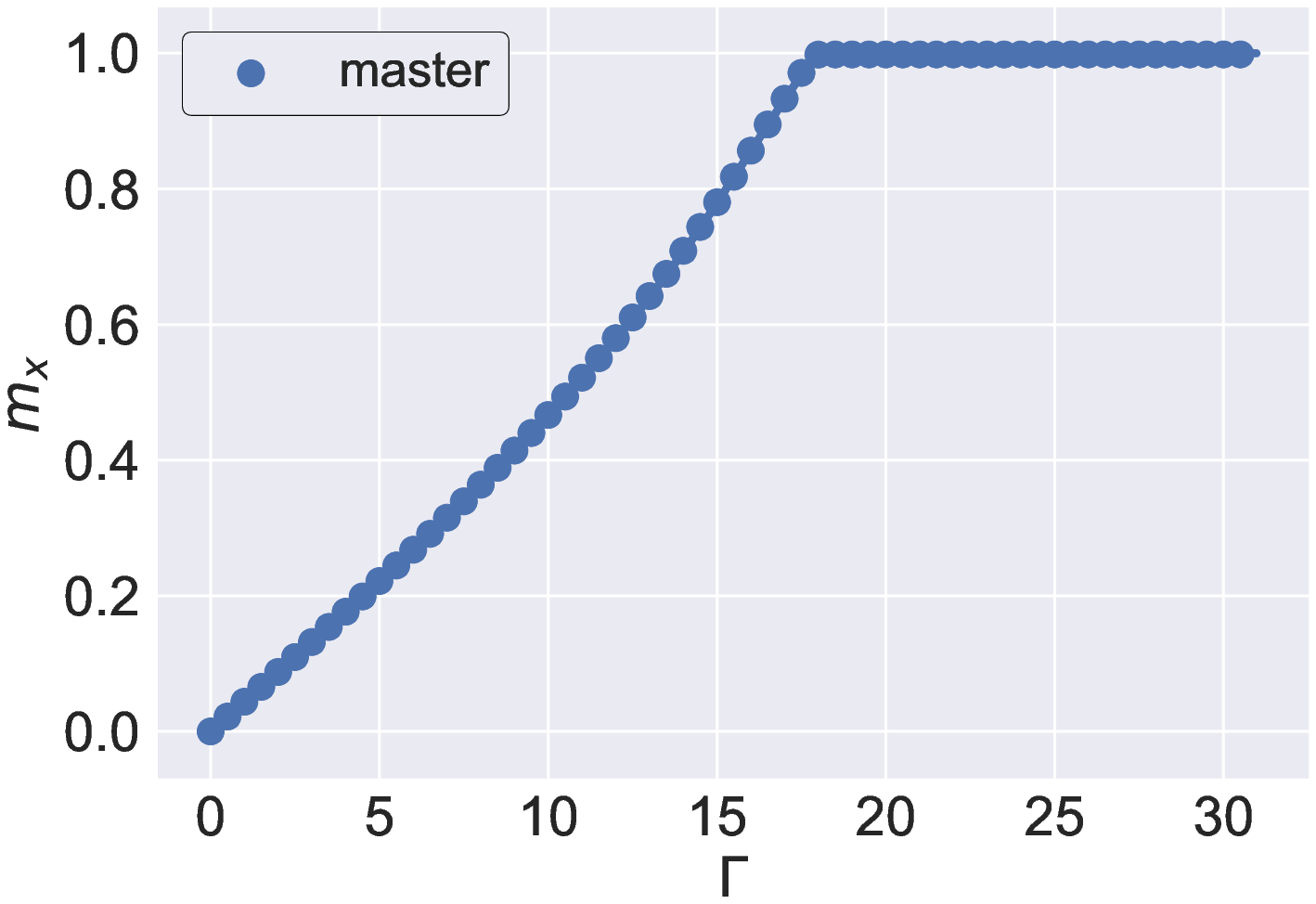}}
\caption{Order parameters with anti-ferromagnetic XX interaction from the master equation and from the saddle-point equations. The figure on the left shows the longitudinal magnetization and that on the right shows the transverse magnetization. The vertical axis denotes these order parameters. The horizontal axis denotes the strength of the transverse magnetic field. The circle and the solid line denote what they do in Fig. \ref{fig:fig_4}.}
\label{fig:fig_8} 
\end{figure*}

\begin{figure*}[t]
\centering
\subfigure[ \label{fig:fig_9a} With no metastable solutions in $\gamma=0$]{\includegraphics[width=50mm]{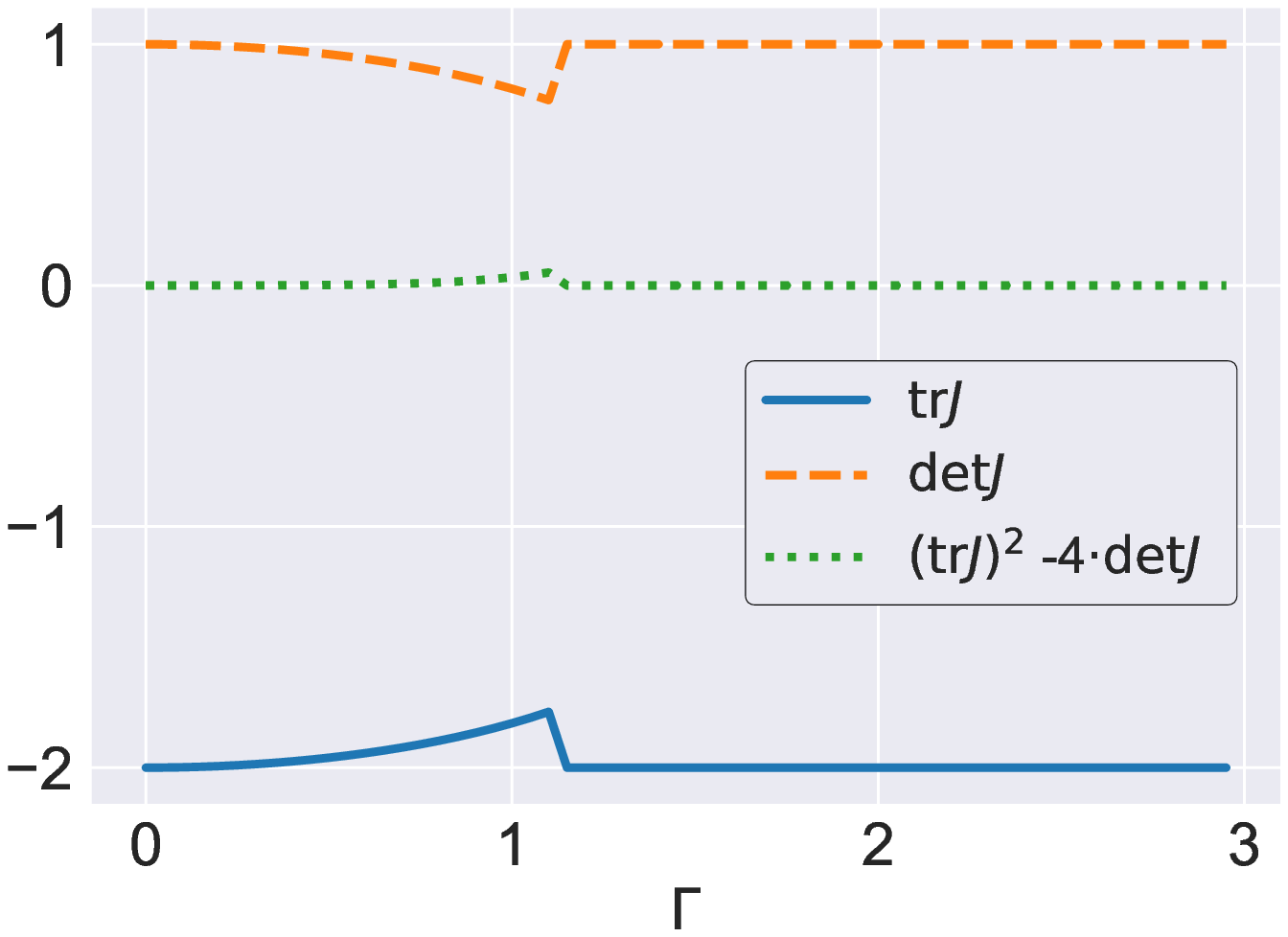}}
\subfigure[\label{fig:fig_9b} With metastable solutions in $\gamma=0$]{\includegraphics[width=50mm]{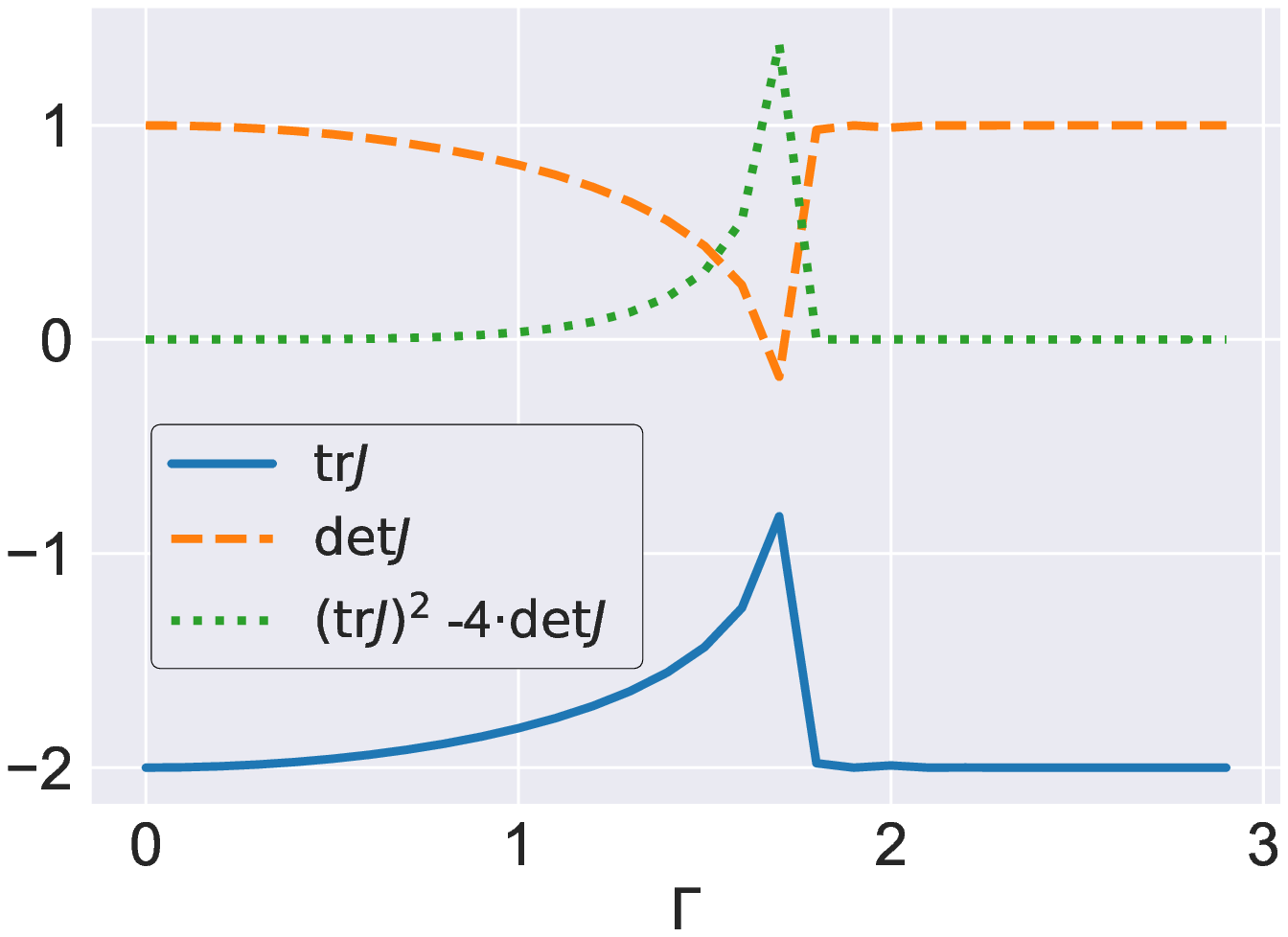}}
\subfigure[ \label{fig:fig_9c} $\gamma=18$]{\includegraphics[width=50mm]{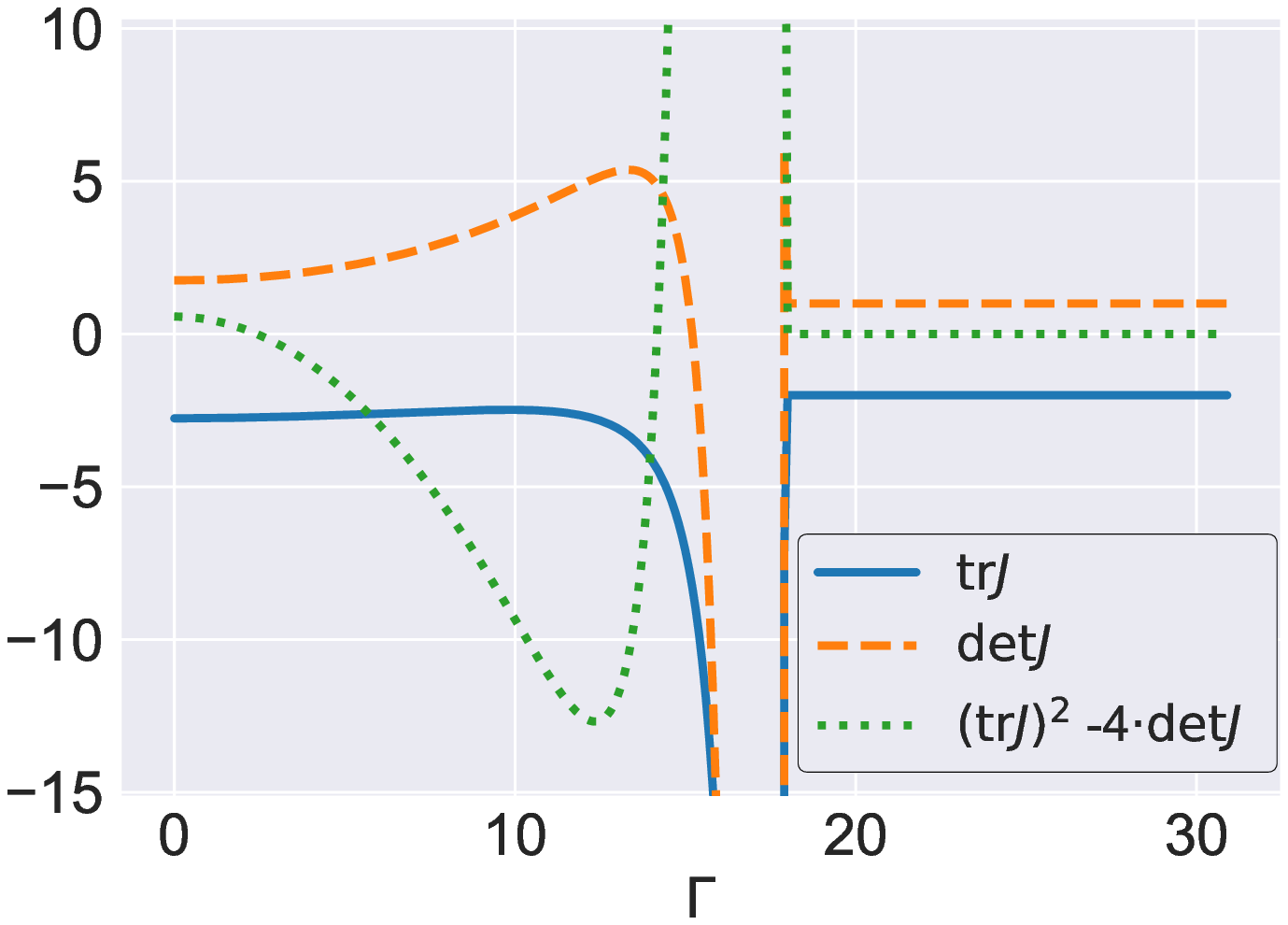}}
\caption{Trace of Jacobian matrix, determinant, and condition with real roots of eigenvalues utilizing  the solutions $(m^*,m_x^*)$ of Eqs. (\ref{eq24}) and (\ref{eq25}). The horizontal axis denotes the strength of the transverse magnetic field. In Figs.  \ref{fig:fig_9a} and \ref{fig:fig_9b} without anti-ferromagnetic XX interaction $\gamma=0$, the solutions do and do not include the metastable solutions, respectively. In Fig. \ref{fig:fig_9c}, we consider the case with anti-ferromagnetic XX interaction $\gamma=18$.
  } \label{fig:fig_9} 
\end{figure*}

 The dynamics of these order parameters with  anti-ferromagnetic XX interaction $\gamma=18$ are shown Figs. \ref{fig:fig_5} and  \ref{fig:fig_6}.
We set three different conditions: $\Gamma=5$ in the ferromagnetic phase, $\Gamma=15$ near the critical point, and $\Gamma=25$ in the paramagnetic phase. These figures indicate that the dynamics exponentially converge to the only steady state. 

 We plot the pseudo free energy in Fig. \ref{fig:fig_7}. The equilibrium solutions obtained by the master equation are consistent with the solutions that minimize the pseudo free energy. The metastable state $m\simeq0$  in $\gamma=0$ vanishes in the equilibrium state. A non-zero solution $m>0$ appears near the critical point. Finally, we can avoid the first-order phase transition. 

To confirm the solution from the master equation, we plot the equilibrium solutions from the master equation with finite inverse temperature $\beta=100$ and the exact solutions from the saddle-point equations in Fig. \ref{fig:fig_8}. From this figure, we can see that the equilibrium solutions from the master equation are consistent with the saddle-point solutions. 

Finally, we investigate the stability of the solutions obtained by the deterministic flow equations. We numerically compute the trace of the Jacobian matrix, the determinant, and the condition with real roots of eigenvalues by utilizing the equilibrium solutions $(m^*,m_x^*)$ of Eqs.  (\ref{eq24}) and (\ref{eq25}). We plot the behavior of the trace of the Jacobian matrix, the determinant, and the condition with real roots of eigenvalues with and without anti-ferromagnetic XX interaction in Fig. \ref{fig:fig_9}. Figure \ref{fig:fig_9a} shows that the equilibrium solutions are stable nodes when the equilibrium solutions are minimum values of the free energy because $(\mathrm{tr} J)^2>4\cdot \mathrm{det} J$, $\mathrm{det} J>0$, and $\mathrm{tr} J<0$ hold. We can see the differences between Fig. \ref{fig:fig_9a} and Fig. \ref{fig:fig_9b}. If the  solutions are not necessarily minimum values of the free energy, the solutions are saddled and unstable near the spinodal point. Then, $(\mathrm{tr} J)^2>4\cdot\mathrm{det} J$, $\mathrm{det} J<0$, and $\mathrm{tr} J<0$ are established. 
In the case of a non-stoquasitc Hamiltonian $\gamma=18$ shown in Fig. \ref{fig:fig_9c}, the equilibrium solutions are stable nodes  in the ferromagnetic phase and the paramagnetic phase. Near the point $\Gamma=\gamma m_x$, the solutions are saddled and unstable. Subsequently, the equilibrium solutions get stable again. As the strength of the transverse magnetic field decreases, the equilibrium solutions become a stable spiral and are asymptotically stable because  $(\mathrm{tr} J)^2<4\cdot \mathrm{det} J$, $\mathrm{det} J>0$ and $\mathrm{tr} J<0$ hold .  In this region, there is an oscillation because the eigenvalues of the Jacobian matrix have complex values. Thus, strong quantum fluctuations like the anti-ferromagnetic XX interaction affect the stability of the equilibrium solutions. 

\section{\label{sec:sec6}Conclusion}

We derived macroscopically deterministic flow equations of order parameters from the Glauber-type master equation under the Suzuki--Trotter decomposition. In AQMC, the adaptive transverse magnetic field is governed by transverse magnetization. By changing the adaptive transverse magnetic field in accordance with the saddle-point solution, we can obtain the dynamics of order parameters of the $p$-spin model with and without anti-ferromagnetic XX interaction. We found that the equilibrium solutions obtained by the deterministic flow equations are identical to the saddle-point solutions obtained by mean-field theory. We can obtain the behavior of order parameters until the system is equilibrated. Without anti-ferromagnetic XX interaction, the metastable state appeared near the spinodal point because the original model has a first-order phase transition. After the spinodal point, the equilibrium solutions converged to different values depending on the initial conditions, owing to the existence of the metastable solution. 
By adding the anti-ferromagnetic XX interaction, the metastable solution vanished. Therefore, the equilibrium solutions converged to the saddle-point solution in all phases. We confirmed that the equilibrium solutions minimize the free energy. Finally, we investigated the stability of the  equilibrium solutions  under a zero-temperature limit. If  the equilibrium solutions include metastable solutions in the case without anti-ferromagnetic XX interaction, the stability of the solutions are unstable. For the model  with anti-ferromagnetic XX interaction, the stability of the equilibrium solutions significantly changed compared to the original model. We found that these strong quantum fluctuations  have an impact on the stability of the equilibrium solutions. 

This approach to use the master equation is useful for understanding the dynamics of QA, which is inhomogeneous the Markovian stochastic process where the strength of the transverse magnetic field is time-dependent. This helps us to investigate not only conventional QA  but also QA with a non-stoquastic Hamiltonian, inhomogeneous driving of the transverse field, and  reverse annealing \cite{reverse_anneal,inhomo,RA_ohkuwa,susa_inhomo}. In our future work, we will analyze the dynamics of these new types of QA.

\section*{Acknowledgments}
M.O. was supported by KAKENHI No. 15H0369, No. 16H04382, and No. 16K13849, ImPACT, and JST-START. 
K.T. was supported by JSPS KAKENHI (No.18H03303). This work was partly supported by JST-CREST (No.JPMJCR1402). 
\appendix
\section{Derivation of differential equations of order parameters}
We here drive the differential equations (\ref{dy_mz}) and (\ref{dy_mx}). 
To derive the deterministic flow equations of order parameters, we utilize the following assumptions:
\begin{align}
&\lim_{N\rightarrow \infty}\frac{1}{N}\sum_{i=1}^N\tanh(\beta\Phi_i(\bm{\sigma}_k:\sigma_{ik\pm1}))
=\left\langle \tanh(\beta\Phi(k))\right\rangle_{\backslash \sigma_k}, \\
&\lim_{N\rightarrow \infty}\frac{K}{N}\sum_{i=1}^N\sigma_{ik+1}\tanh(\beta\Phi_i(\bm{\sigma}_k):\sigma_{ik\pm1})=\left\langle K\sigma_{k+1}\tanh(\beta\Phi(k))\right\rangle_{\backslash \sigma_k}
\end{align}
where we define the effective single local field $\beta\Phi(k)\equiv (\beta p/M)m_k^{p-1}+B/2(\sigma_{k-1}+\sigma_{k+1})$ on the $k$-th Trotter slice and the average 
\begin{align}
\langle \cdots \rangle_{\backslash \sigma_k} &\equiv
\lim_{M\rightarrow \infty}\frac{ \sum_{\bm{\sigma} \backslash \sigma_k} (\cdots)\exp(\beta \sum_{l\neq k}^M\Phi(l)\sigma_{l})}{ \sum_{\bm{\sigma} \backslash \sigma_k }\exp(\beta \sum_{l\neq k}^M\Phi(l)\sigma_{l})}.
\end{align}
Here, we write the summation with respect to all sites  except for the Trotter slice  $k$ as 
\begin{align}
 \sum_{\bm{\sigma} \backslash \sigma_k }\equiv\sum_{\sigma_1} \cdots \sum_{\sigma_{k-1}}  \sum_{\sigma_{k+1}} \cdots \sum_{\sigma_{M}}.
\end{align}
We can rewrite  $\tanh(\beta \Phi(k))$ and $\sigma_{k+1}\tanh(\beta \Phi(k))$ as 
\begin{align}
\tanh(\beta\Phi(k))&=\frac{\sum_{\sigma_{k}=\pm1}\sigma_{k} \exp(\beta \Phi(k)\sigma_{k})}{\sum_{\sigma_{k}=\pm1}\exp(\beta \Phi(k)\sigma_{k})} ,\nonumber \\
\end{align}
\begin{align}
\sigma_{k+1}\tanh(\beta\Phi(k))&=\frac{\sum_{\sigma_{k}=\pm1}\sigma_{k}\sigma_{k+1}\exp(\beta \Phi(k)\sigma_{k})}{\sum_{\sigma_{k}=\pm1}\exp(\beta \Phi(k)\sigma_{k})}.
\end{align}
In a manner similar to a previous study \cite{determ_order_qmc}, we can obtain the expectation of $\tanh(\beta \Phi(k))$ and that of $K\sigma_{k+1}\tanh(\beta \Phi(k))$ as
\begin{align}
&\langle \tanh(\beta\Phi(k))\rangle_{\backslash \sigma_k}\nonumber \\
&=\lim_{M\rightarrow \infty}\frac{ \sum_{\bm{\sigma} } \sigma_k \exp(\beta \sum_{l=1}^M\Phi(l)\sigma_{l})}{ \sum_{\bm{\sigma}  }\exp(\beta \sum_{l=1}^M\Phi(l)\sigma_{l})}\nonumber \\
&\equiv\langle \sigma_k \rangle_{path}\label{a_7},
\end{align}
\begin{align}
&\langle K\sigma_{k+1}\tanh(\beta\Phi(k))\rangle_{\backslash \sigma_k}\nonumber \\
&=\lim_{M\rightarrow \infty}\frac{ \sum_{\bm{\sigma} } K\sigma_k\sigma_{k+1} \exp(\beta \sum_{l=1}^M\Phi(l)\sigma_{l})}{ \sum_{\bm{\sigma}  }\exp(\beta \sum_{l=1}^M\Phi(l)\sigma_{l})}\nonumber \\
&\equiv \langle K\sigma_k\sigma_{k+1}\rangle_{path}\label{a_8}.
\end{align}
We substitute Eqs. (\ref{a_7}) and (\ref{a_8})  for Eq. (\ref{order_mastert}). The differential equations are written as 
\begin{align}
&\frac{dP_{t}(\left\{m_k\right\},\left\{m_{xk}\right\})}{dt}=\sum_{k} \frac{\partial  }{\partial m_k}m_k P_{t}(\left\{m_k\right\},\left\{m_{xk}\right\})\nonumber \\
&- \sum_{k}\frac{\partial  }{\partial m_k}\left\{P_{t}(\left\{m_k\right\},\left\{m_{xk}\right\}) \langle \sigma_k\rangle_{path}\right\} \nonumber \\
&+\sum_{k}\frac{\partial  }{\partial m_{xk}}m_{xk} P_{t}(\left\{m_k\right\},\left\{m_{xk}\right\})\nonumber \\
&-\sum_{k} \frac{\partial  }{\partial m_{xk}}\left\{P_{t}(\left\{m_k\right\},\left\{m_{xk}\right\}) \langle K\sigma_k\sigma_{k+1}\rangle_{path}\right\} .\label{order_mastert_nsa}
\end{align}
In order to derive a compact representation of the differential equations, we substitute $P_{t}(\left\{m_k\right\},\left\{m_{xk}\right\})=\prod_{k=1}^M\delta(m_k-m_k(t))\delta(m_{xk}-m_{xk}(t))$ into Eq. (\ref{order_mastert_nsa}) and carry out the integral with respect to $\prod_{k}m_k$ and $\prod_{k}m_{xk}$  after multiplying itself $m_k$. Finally, we can obtain the differential equations for each Trotter slices $k$ as
\begin{align}
\frac{dm_k}{dt}=-m_k+\langle \sigma_k\rangle_{path}\label{eq10}.
\end{align}

In order to derive a compact representation of the differential equation, 
we utilize the static approximation $m_k=m, m_{xk}=m_x$. Under this approximation, we inverse the procedure of the Suzuki--Trotter decomposition:
\begin{align}
Z(\tilde{m}_x)&= \lim_{M\rightarrow \infty} \sum_{\bm{\sigma}}  \exp \left(\frac{\beta p m^{p-1}}{M}\sum_{k}\sigma_k + B\sum_k \sigma_k \sigma_{k+1 }\right)  \nonumber \\
&\propto \mathrm{Tr} \Biggl\{ \exp\left(\beta pm^{p-1} \sigma^z+\beta \tilde{m}_x \sigma^x\right) \Biggr\}\nonumber \\
&=2\cosh\left(\beta \sqrt{\left(pm^{p-1}\right)^2+\left(\tilde{m}_x\right)^2}\right).
 \end{align}
We can regard $\langle \sigma_k \rangle_{path}=\lim_{M\rightarrow \infty}\langle M^{-1}\sum_k \sigma_k\rangle_{path}$ as 
\begin{align}
&\langle \sigma_k \rangle_{path}\nonumber \\
&=\lim_{M\rightarrow \infty}  \frac{ \sum_{\bm{\sigma} } M^{-1}\sum_k \sigma_k\exp \left(\frac{\beta pm^{p-1}}{M}\sum_{k}\sigma_k + B\sum_k \sigma_k \sigma_{k+1 }\right)}{ \sum_{\bm{\sigma} } \exp \left(\frac{\beta pm^{p-1}}{M}\sum_{k}\sigma_k + B\sum_k \sigma_k \sigma_{k+1 }\right)}\nonumber \\
&=\lim_{M\rightarrow \infty} \frac{\partial \log Z(\tilde{m}_x)}{\partial (\beta pm^{p-1})}\nonumber \\
&\propto\frac{pm^{p-1}}{\sqrt{(pm^{p-1})^2+\tilde{m}_x^2}}\tanh\left(\beta \sqrt{(pm^{p-1})^2+\tilde{m}_x^2}\right).
\label{path_m}
\end{align}
We substitute Eq. (\ref{path_m}) for Eq. (\ref{eq10}) and obtain the deterministic equation (\ref{dy_mz}).
\vspace{8pt}
For $m_{xk}$, we similarly consider the flow equation as 
\begin{align}
\frac{dm_{xk}}{dt}=-m_{xk}+\langle K\sigma_k\sigma_{k+1}\rangle_{path}.
\label{col_dy}
\end{align}
Under the static approximation, we have  $\bigl<  K\sigma_k \sigma_{k+1} \bigr> _{path}=\lim_{M\rightarrow \infty}\langle M^{-1}\sum_k K\sigma_k \sigma_{k+1}\rangle_{path}$ as 
\begin{align}
&\langle\frac{K}{M} \sum_{k=1}^M\sigma_k \sigma_{k+1} \rangle_{path}\nonumber \\
&=\lim_{M\rightarrow \infty}  \frac{\mathrm{Tr}M^{-1}K\sum_k \sigma_k \sigma_{k+1}\exp \left(\frac{\beta pm^{p-1}}{M}\sum_{k}\sigma_k + B\sum_k \sigma_k \sigma_{k+1 }\right)}{\mathrm{Tr}\exp \left(\frac{\beta pm^{p-1}}{M}\sum_{k}\sigma_k + B\sum_k \sigma_k \sigma_{k+1 }\right)}\nonumber \\
&=\lim_{M\rightarrow \infty} \frac{\partial \log Z(\tilde{m}_x)}{\partial (\beta \tilde{m}_x )} \nonumber \\
&\propto\frac{\tilde{m}_x}{\sqrt{(pm^{p-1})^2+\tilde{m}_x^2}}\tanh\left(\beta \sqrt{(pm^{p-1})^2+\tilde{m}_x^2}\right).
\label{col_dy3}
\end{align}
After assigning Eq. (\ref{col_dy3}) to Eq. (\ref{col_dy}), we can obtain Eq. (\ref{dy_mx}).

\bibliography{main.bib}

\begin{thebibliography}{54}%
\makeatletter
\providecommand \@ifxundefined [1]{%
 \@ifx{#1\undefined}
}%
\providecommand \@ifnum [1]{%
 \ifnum #1\expandafter \@firstoftwo
 \else \expandafter \@secondoftwo
 \fi
}%
\providecommand \@ifx [1]{%
 \ifx #1\expandafter \@firstoftwo
 \else \expandafter \@secondoftwo
 \fi
}%
\providecommand \natexlab [1]{#1}%
\providecommand \enquote  [1]{``#1''}%
\providecommand \bibnamefont  [1]{#1}%
\providecommand \bibfnamefont [1]{#1}%
\providecommand \citenamefont [1]{#1}%
\providecommand \href@noop [0]{\@secondoftwo}%
\providecommand \href [0]{\begingroup \@sanitize@url \@href}%
\providecommand \@href[1]{\@@startlink{#1}\@@href}%
\providecommand \@@href[1]{\endgroup#1\@@endlink}%
\providecommand \@sanitize@url [0]{\catcode `\\12\catcode `\$12\catcode
  `\&12\catcode `\#12\catcode `\^12\catcode `\_12\catcode `\%12\relax}%
\providecommand \@@startlink[1]{}%
\providecommand \@@endlink[0]{}%
\providecommand \url  [0]{\begingroup\@sanitize@url \@url }%
\providecommand \@url [1]{\endgroup\@href {#1}{\urlprefix }}%
\providecommand \urlprefix  [0]{URL }%
\providecommand \Eprint [0]{\href }%
\providecommand \doibase [0]{http://dx.doi.org/}%
\providecommand \selectlanguage [0]{\@gobble}%
\providecommand \bibinfo  [0]{\@secondoftwo}%
\providecommand \bibfield  [0]{\@secondoftwo}%
\providecommand \translation [1]{[#1]}%
\providecommand \BibitemOpen [0]{}%
\providecommand \bibitemStop [0]{}%
\providecommand \bibitemNoStop [0]{.\EOS\space}%
\providecommand \EOS [0]{\spacefactor3000\relax}%
\providecommand \BibitemShut  [1]{\csname bibitem#1\endcsname}%
\let\auto@bib@innerbib\@empty
\bibitem [{\citenamefont {Kadowaki}\ and\ \citenamefont
  {Nishimori}(1998)}]{kadowaki_nishimori}%
  \BibitemOpen
  \bibfield  {author} {\bibinfo {author} {\bibfnamefont {T.}~\bibnamefont
  {Kadowaki}}\ and\ \bibinfo {author} {\bibfnamefont {H.}~\bibnamefont
  {Nishimori}},\ }\href {\doibase 10.1103/PhysRevE.58.5355} {\bibfield
  {journal} {\bibinfo  {journal} {Phys. Rev. E}\ }\textbf {\bibinfo {volume}
  {58}},\ \bibinfo {pages} {5355} (\bibinfo {year} {1998})}\BibitemShut
  {NoStop}%
\bibitem [{\citenamefont {Das}\ and\ \citenamefont {Chakrabarti}(2008)}]{qa2}%
  \BibitemOpen
  \bibfield  {author} {\bibinfo {author} {\bibfnamefont {A.}~\bibnamefont
  {Das}}\ and\ \bibinfo {author} {\bibfnamefont {B.~K.}\ \bibnamefont
  {Chakrabarti}},\ }\href {\doibase 10.1103/RevModPhys.80.1061} {\bibfield
  {journal} {\bibinfo  {journal} {Rev. Mod. Phys.}\ }\textbf {\bibinfo {volume}
  {80}},\ \bibinfo {pages} {1061} (\bibinfo {year} {2008})}\BibitemShut
  {NoStop}%
\bibitem [{\citenamefont {Johnson}\ \emph {et~al.}(2010)\citenamefont
  {Johnson}, \citenamefont {Bunyk}, \citenamefont {Maibaum}, \citenamefont
  {Tolkacheva}, \citenamefont {Berkley}, \citenamefont {Chapple}, \citenamefont
  {Harris}, \citenamefont {Johansson}, \citenamefont {Lanting}, \citenamefont
  {Perminov}, \citenamefont {Ladizinsky}, \citenamefont {Oh},\ and\
  \citenamefont {Rose}}]{Dwave2010a}%
  \BibitemOpen
  \bibfield  {author} {\bibinfo {author} {\bibfnamefont {M.~W.}\ \bibnamefont
  {Johnson}}, \bibinfo {author} {\bibfnamefont {P.}~\bibnamefont {Bunyk}},
  \bibinfo {author} {\bibfnamefont {F.}~\bibnamefont {Maibaum}}, \bibinfo
  {author} {\bibfnamefont {E.}~\bibnamefont {Tolkacheva}}, \bibinfo {author}
  {\bibfnamefont {A.~J.}\ \bibnamefont {Berkley}}, \bibinfo {author}
  {\bibfnamefont {E.~M.}\ \bibnamefont {Chapple}}, \bibinfo {author}
  {\bibfnamefont {R.}~\bibnamefont {Harris}}, \bibinfo {author} {\bibfnamefont
  {J.}~\bibnamefont {Johansson}}, \bibinfo {author} {\bibfnamefont
  {T.}~\bibnamefont {Lanting}}, \bibinfo {author} {\bibfnamefont
  {I.}~\bibnamefont {Perminov}}, \bibinfo {author} {\bibfnamefont
  {E.}~\bibnamefont {Ladizinsky}}, \bibinfo {author} {\bibfnamefont
  {T.}~\bibnamefont {Oh}}, \ and\ \bibinfo {author} {\bibfnamefont
  {G.}~\bibnamefont {Rose}},\ }\href@noop {} {\bibfield  {journal} {\bibinfo
  {journal} {Superconductor Science and Technology}\ }\textbf {\bibinfo
  {volume} {23}},\ \bibinfo {pages} {065004} (\bibinfo {year}
  {2010})}\BibitemShut {NoStop}%
\bibitem [{\citenamefont {Berkley}\ \emph {et~al.}(2010)\citenamefont
  {Berkley}, \citenamefont {Johnson}, \citenamefont {Bunyk}, \citenamefont
  {Harris}, \citenamefont {Johansson}, \citenamefont {Lanting}, \citenamefont
  {Ladizinsky}, \citenamefont {Tolkacheva}, \citenamefont {Amin},\ and\
  \citenamefont {Rose}}]{Dwave2010b}%
  \BibitemOpen
  \bibfield  {author} {\bibinfo {author} {\bibfnamefont {A.~J.}\ \bibnamefont
  {Berkley}}, \bibinfo {author} {\bibfnamefont {M.~W.}\ \bibnamefont
  {Johnson}}, \bibinfo {author} {\bibfnamefont {P.}~\bibnamefont {Bunyk}},
  \bibinfo {author} {\bibfnamefont {R.}~\bibnamefont {Harris}}, \bibinfo
  {author} {\bibfnamefont {J.}~\bibnamefont {Johansson}}, \bibinfo {author}
  {\bibfnamefont {T.}~\bibnamefont {Lanting}}, \bibinfo {author} {\bibfnamefont
  {E.}~\bibnamefont {Ladizinsky}}, \bibinfo {author} {\bibfnamefont
  {E.}~\bibnamefont {Tolkacheva}}, \bibinfo {author} {\bibfnamefont {M.~H.~S.}\
  \bibnamefont {Amin}}, \ and\ \bibinfo {author} {\bibfnamefont
  {G.}~\bibnamefont {Rose}},\ }\href@noop {} {\bibfield  {journal} {\bibinfo
  {journal} {Superconductor Science and Technology}\ }\textbf {\bibinfo
  {volume} {23}},\ \bibinfo {pages} {105014} (\bibinfo {year}
  {2010})}\BibitemShut {NoStop}%
\bibitem [{\citenamefont {Harris}\ \emph {et~al.}(2010)\citenamefont {Harris},
  \citenamefont {Johnson}, \citenamefont {Lanting}, \citenamefont {Berkley},
  \citenamefont {Johansson}, \citenamefont {Bunyk}, \citenamefont {Tolkacheva},
  \citenamefont {Ladizinsky}, \citenamefont {Ladizinsky}, \citenamefont {Oh},
  \citenamefont {Cioata}, \citenamefont {Perminov}, \citenamefont {Spear},
  \citenamefont {Enderud}, \citenamefont {Rich}, \citenamefont {Uchaikin},
  \citenamefont {Thom}, \citenamefont {Chapple}, \citenamefont {Wang},
  \citenamefont {Wilson}, \citenamefont {Amin}, \citenamefont {Dickson},
  \citenamefont {Karimi}, \citenamefont {Macready}, \citenamefont {Truncik},\
  and\ \citenamefont {Rose}}]{Dwave2010c}%
  \BibitemOpen
  \bibfield  {author} {\bibinfo {author} {\bibfnamefont {R.}~\bibnamefont
  {Harris}}, \bibinfo {author} {\bibfnamefont {M.~W.}\ \bibnamefont {Johnson}},
  \bibinfo {author} {\bibfnamefont {T.}~\bibnamefont {Lanting}}, \bibinfo
  {author} {\bibfnamefont {A.~J.}\ \bibnamefont {Berkley}}, \bibinfo {author}
  {\bibfnamefont {J.}~\bibnamefont {Johansson}}, \bibinfo {author}
  {\bibfnamefont {P.}~\bibnamefont {Bunyk}}, \bibinfo {author} {\bibfnamefont
  {E.}~\bibnamefont {Tolkacheva}}, \bibinfo {author} {\bibfnamefont
  {E.}~\bibnamefont {Ladizinsky}}, \bibinfo {author} {\bibfnamefont
  {N.}~\bibnamefont {Ladizinsky}}, \bibinfo {author} {\bibfnamefont
  {T.}~\bibnamefont {Oh}}, \bibinfo {author} {\bibfnamefont {F.}~\bibnamefont
  {Cioata}}, \bibinfo {author} {\bibfnamefont {I.}~\bibnamefont {Perminov}},
  \bibinfo {author} {\bibfnamefont {P.}~\bibnamefont {Spear}}, \bibinfo
  {author} {\bibfnamefont {C.}~\bibnamefont {Enderud}}, \bibinfo {author}
  {\bibfnamefont {C.}~\bibnamefont {Rich}}, \bibinfo {author} {\bibfnamefont
  {S.}~\bibnamefont {Uchaikin}}, \bibinfo {author} {\bibfnamefont {M.~C.}\
  \bibnamefont {Thom}}, \bibinfo {author} {\bibfnamefont {E.~M.}\ \bibnamefont
  {Chapple}}, \bibinfo {author} {\bibfnamefont {J.}~\bibnamefont {Wang}},
  \bibinfo {author} {\bibfnamefont {B.}~\bibnamefont {Wilson}}, \bibinfo
  {author} {\bibfnamefont {M.~H.~S.}\ \bibnamefont {Amin}}, \bibinfo {author}
  {\bibfnamefont {N.}~\bibnamefont {Dickson}}, \bibinfo {author} {\bibfnamefont
  {K.}~\bibnamefont {Karimi}}, \bibinfo {author} {\bibfnamefont
  {B.}~\bibnamefont {Macready}}, \bibinfo {author} {\bibfnamefont {C.~J.~S.}\
  \bibnamefont {Truncik}}, \ and\ \bibinfo {author} {\bibfnamefont
  {G.}~\bibnamefont {Rose}},\ }\href {\doibase 10.1103/PhysRevB.82.024511}
  {\bibfield  {journal} {\bibinfo  {journal} {Phys. Rev. B}\ }\textbf {\bibinfo
  {volume} {82}},\ \bibinfo {pages} {024511} (\bibinfo {year}
  {2010})}\BibitemShut {NoStop}%
\bibitem [{\citenamefont {Johnson}\ \emph {et~al.}(2011)\citenamefont
  {Johnson}, \citenamefont {Amin}, \citenamefont {Gildert}, \citenamefont
  {Lanting}, \citenamefont {Hamze}, \citenamefont {Dickson}, \citenamefont
  {Harris}, \citenamefont {Berkley}, \citenamefont {Johansson}, \citenamefont
  {Bunyk}, \citenamefont {Chapple}, \citenamefont {Enderud}, \citenamefont
  {Hilton}, \citenamefont {Karimi}, \citenamefont {Ladizinsky}, \citenamefont
  {Ladizinsky}, \citenamefont {Oh}, \citenamefont {Perminov}, \citenamefont
  {Rich}, \citenamefont {Thom}, \citenamefont {Tolkacheva}, \citenamefont
  {Truncik}, \citenamefont {Uchaikin}, \citenamefont {Wang}, \citenamefont
  {Wilson},\ and\ \citenamefont {Rose}}]{dwave_machine}%
  \BibitemOpen
  \bibfield  {author} {\bibinfo {author} {\bibfnamefont {M.~W.}\ \bibnamefont
  {Johnson}}, \bibinfo {author} {\bibfnamefont {M.~H.~S.}\ \bibnamefont
  {Amin}}, \bibinfo {author} {\bibfnamefont {S.}~\bibnamefont {Gildert}},
  \bibinfo {author} {\bibfnamefont {T.}~\bibnamefont {Lanting}}, \bibinfo
  {author} {\bibfnamefont {F.}~\bibnamefont {Hamze}}, \bibinfo {author}
  {\bibfnamefont {N.}~\bibnamefont {Dickson}}, \bibinfo {author} {\bibfnamefont
  {R.}~\bibnamefont {Harris}}, \bibinfo {author} {\bibfnamefont {A.~J.}\
  \bibnamefont {Berkley}}, \bibinfo {author} {\bibfnamefont {J.}~\bibnamefont
  {Johansson}}, \bibinfo {author} {\bibfnamefont {P.}~\bibnamefont {Bunyk}},
  \bibinfo {author} {\bibfnamefont {E.~M.}\ \bibnamefont {Chapple}}, \bibinfo
  {author} {\bibfnamefont {C.}~\bibnamefont {Enderud}}, \bibinfo {author}
  {\bibfnamefont {J.~P.}\ \bibnamefont {Hilton}}, \bibinfo {author}
  {\bibfnamefont {K.}~\bibnamefont {Karimi}}, \bibinfo {author} {\bibfnamefont
  {E.}~\bibnamefont {Ladizinsky}}, \bibinfo {author} {\bibfnamefont
  {N.}~\bibnamefont {Ladizinsky}}, \bibinfo {author} {\bibfnamefont
  {T.}~\bibnamefont {Oh}}, \bibinfo {author} {\bibfnamefont {I.}~\bibnamefont
  {Perminov}}, \bibinfo {author} {\bibfnamefont {C.}~\bibnamefont {Rich}},
  \bibinfo {author} {\bibfnamefont {M.~C.}\ \bibnamefont {Thom}}, \bibinfo
  {author} {\bibfnamefont {E.}~\bibnamefont {Tolkacheva}}, \bibinfo {author}
  {\bibfnamefont {C.~J.~S.}\ \bibnamefont {Truncik}}, \bibinfo {author}
  {\bibfnamefont {S.}~\bibnamefont {Uchaikin}}, \bibinfo {author}
  {\bibfnamefont {J.}~\bibnamefont {Wang}}, \bibinfo {author} {\bibfnamefont
  {B.}~\bibnamefont {Wilson}}, \ and\ \bibinfo {author} {\bibfnamefont
  {G.}~\bibnamefont {Rose}},\ }\href {http://dx.doi.org/10.1038/nature10012}
  {\bibfield  {journal} {\bibinfo  {journal} {Nature}\ }\textbf {\bibinfo
  {volume} {473}},\ \bibinfo {pages} {194 EP } (\bibinfo {year}
  {2011})}\BibitemShut {NoStop}%
\bibitem [{\citenamefont {Bunyk}\ \emph {et~al.}(2014)\citenamefont {Bunyk},
  \citenamefont {Hoskinson}, \citenamefont {Johnson}, \citenamefont
  {Tolkacheva}, \citenamefont {Altomare}, \citenamefont {Berkley},
  \citenamefont {Harris}, \citenamefont {Hilton}, \citenamefont {Lanting},
  \citenamefont {Przybysz},\ and\ \citenamefont {Whittaker}}]{Dwave2014}%
  \BibitemOpen
  \bibfield  {author} {\bibinfo {author} {\bibfnamefont {P.~I.}\ \bibnamefont
  {Bunyk}}, \bibinfo {author} {\bibfnamefont {E.~M.}\ \bibnamefont
  {Hoskinson}}, \bibinfo {author} {\bibfnamefont {M.~W.}\ \bibnamefont
  {Johnson}}, \bibinfo {author} {\bibfnamefont {E.}~\bibnamefont {Tolkacheva}},
  \bibinfo {author} {\bibfnamefont {F.}~\bibnamefont {Altomare}}, \bibinfo
  {author} {\bibfnamefont {A.~J.}\ \bibnamefont {Berkley}}, \bibinfo {author}
  {\bibfnamefont {R.}~\bibnamefont {Harris}}, \bibinfo {author} {\bibfnamefont
  {J.~P.}\ \bibnamefont {Hilton}}, \bibinfo {author} {\bibfnamefont
  {T.}~\bibnamefont {Lanting}}, \bibinfo {author} {\bibfnamefont {A.~J.}\
  \bibnamefont {Przybysz}}, \ and\ \bibinfo {author} {\bibfnamefont
  {J.}~\bibnamefont {Whittaker}},\ }\href {\doibase 10.1109/TASC.2014.2318294}
  {\bibfield  {journal} {\bibinfo  {journal} {IEEE Transactions on Applied
  Superconductivity}\ }\textbf {\bibinfo {volume} {24}},\ \bibinfo {pages} {1}
  (\bibinfo {year} {2014})}\BibitemShut {NoStop}%
\bibitem [{\citenamefont {Boixo}\ \emph {et~al.}(2014)\citenamefont {Boixo},
  \citenamefont {R{\o}nnow}, \citenamefont {Isakov}, \citenamefont {Wang},
  \citenamefont {Wecker}, \citenamefont {Lidar}, \citenamefont {Martinis},\
  and\ \citenamefont {Troyer}}]{dwave3}%
  \BibitemOpen
  \bibfield  {author} {\bibinfo {author} {\bibfnamefont {S.}~\bibnamefont
  {Boixo}}, \bibinfo {author} {\bibfnamefont {T.~F.}\ \bibnamefont
  {R{\o}nnow}}, \bibinfo {author} {\bibfnamefont {S.~V.}\ \bibnamefont
  {Isakov}}, \bibinfo {author} {\bibfnamefont {Z.}~\bibnamefont {Wang}},
  \bibinfo {author} {\bibfnamefont {D.}~\bibnamefont {Wecker}}, \bibinfo
  {author} {\bibfnamefont {D.~A.}\ \bibnamefont {Lidar}}, \bibinfo {author}
  {\bibfnamefont {J.~M.}\ \bibnamefont {Martinis}}, \ and\ \bibinfo {author}
  {\bibfnamefont {M.}~\bibnamefont {Troyer}},\ }\href
  {http://dx.doi.org/10.1038/nphys2900} {\bibfield  {journal} {\bibinfo
  {journal} {Nature Physics}\ }\textbf {\bibinfo {volume} {10}},\ \bibinfo
  {pages} {218 EP } (\bibinfo {year} {2014})}\BibitemShut {NoStop}%
\bibitem [{\citenamefont {Denchev}\ \emph {et~al.}(2016)\citenamefont
  {Denchev}, \citenamefont {Boixo}, \citenamefont {Isakov}, \citenamefont
  {Ding}, \citenamefont {Babbush}, \citenamefont {Smelyanskiy}, \citenamefont
  {Martinis},\ and\ \citenamefont {Neven}}]{dwave2}%
  \BibitemOpen
  \bibfield  {author} {\bibinfo {author} {\bibfnamefont {V.~S.}\ \bibnamefont
  {Denchev}}, \bibinfo {author} {\bibfnamefont {S.}~\bibnamefont {Boixo}},
  \bibinfo {author} {\bibfnamefont {S.~V.}\ \bibnamefont {Isakov}}, \bibinfo
  {author} {\bibfnamefont {N.}~\bibnamefont {Ding}}, \bibinfo {author}
  {\bibfnamefont {R.}~\bibnamefont {Babbush}}, \bibinfo {author} {\bibfnamefont
  {V.}~\bibnamefont {Smelyanskiy}}, \bibinfo {author} {\bibfnamefont
  {J.}~\bibnamefont {Martinis}}, \ and\ \bibinfo {author} {\bibfnamefont
  {H.}~\bibnamefont {Neven}},\ }\href {\doibase 10.1103/PhysRevX.6.031015}
  {\bibfield  {journal} {\bibinfo  {journal} {Phys. Rev. X}\ }\textbf {\bibinfo
  {volume} {6}},\ \bibinfo {pages} {031015} (\bibinfo {year}
  {2016})}\BibitemShut {NoStop}%
\bibitem [{\citenamefont {Amin}(2015)}]{Amin2015}%
  \BibitemOpen
  \bibfield  {author} {\bibinfo {author} {\bibfnamefont {M.~H.}\ \bibnamefont
  {Amin}},\ }\href@noop {} {\bibfield  {journal} {\bibinfo  {journal} {Phys.
  Rev. A}\ }\textbf {\bibinfo {volume} {92}},\ \bibinfo {pages} {052323}
  (\bibinfo {year} {2015})}\BibitemShut {NoStop}%
\bibitem [{\citenamefont {Rosenberg}\ \emph {et~al.}(2016)\citenamefont
  {Rosenberg}, \citenamefont {Haghnegahdar}, \citenamefont {Goddard},
  \citenamefont {Carr}, \citenamefont {Wu},\ and\ \citenamefont
  {de~Prado}}]{Rosenberg2016}%
  \BibitemOpen
  \bibfield  {author} {\bibinfo {author} {\bibfnamefont {G.}~\bibnamefont
  {Rosenberg}}, \bibinfo {author} {\bibfnamefont {P.}~\bibnamefont
  {Haghnegahdar}}, \bibinfo {author} {\bibfnamefont {P.}~\bibnamefont
  {Goddard}}, \bibinfo {author} {\bibfnamefont {P.}~\bibnamefont {Carr}},
  \bibinfo {author} {\bibfnamefont {K.}~\bibnamefont {Wu}}, \ and\ \bibinfo
  {author} {\bibfnamefont {M.~L.}\ \bibnamefont {de~Prado}},\ }\href {\doibase
  10.1109/JSTSP.2016.2574703} {\bibfield  {journal} {\bibinfo  {journal} {IEEE
  Journal of Selected Topics in Signal Processing}\ }\textbf {\bibinfo {volume}
  {10}},\ \bibinfo {pages} {1053} (\bibinfo {year} {2016})}\BibitemShut
  {NoStop}%
\bibitem [{\citenamefont {Perdomo-Ortiz}\ \emph {et~al.}(2012)\citenamefont
  {Perdomo-Ortiz}, \citenamefont {Dickson}, \citenamefont {Drew-Brook},
  \citenamefont {Rose},\ and\ \citenamefont {Aspuru-Guzik}}]{Perdomo2012}%
  \BibitemOpen
  \bibfield  {author} {\bibinfo {author} {\bibfnamefont {A.}~\bibnamefont
  {Perdomo-Ortiz}}, \bibinfo {author} {\bibfnamefont {N.}~\bibnamefont
  {Dickson}}, \bibinfo {author} {\bibfnamefont {M.}~\bibnamefont {Drew-Brook}},
  \bibinfo {author} {\bibfnamefont {G.}~\bibnamefont {Rose}}, \ and\ \bibinfo
  {author} {\bibfnamefont {A.}~\bibnamefont {Aspuru-Guzik}},\ }\href@noop {}
  {\bibfield  {journal} {\bibinfo  {journal} {Scientific Reports}\ }\textbf
  {\bibinfo {volume} {2}},\ \bibinfo {pages} {571 EP } (\bibinfo {year}
  {2012})}\BibitemShut {NoStop}%
\bibitem [{\citenamefont {Hernandez}\ and\ \citenamefont
  {Aramon}(2017)}]{Hernandez2017}%
  \BibitemOpen
  \bibfield  {author} {\bibinfo {author} {\bibfnamefont {M.}~\bibnamefont
  {Hernandez}}\ and\ \bibinfo {author} {\bibfnamefont {M.}~\bibnamefont
  {Aramon}},\ }\href {\doibase 10.1007/s11128-017-1586-y} {\bibfield  {journal}
  {\bibinfo  {journal} {Quantum Information Processing}\ }\textbf {\bibinfo
  {volume} {16}},\ \bibinfo {pages} {133} (\bibinfo {year} {2017})}\BibitemShut
  {NoStop}%
\bibitem [{\citenamefont {Li}\ \emph {et~al.}(2018)\citenamefont {Li},
  \citenamefont {Di~Felice}, \citenamefont {Rohs},\ and\ \citenamefont
  {Lidar}}]{Richard2018}%
  \BibitemOpen
  \bibfield  {author} {\bibinfo {author} {\bibfnamefont {R.~Y.}\ \bibnamefont
  {Li}}, \bibinfo {author} {\bibfnamefont {R.}~\bibnamefont {Di~Felice}},
  \bibinfo {author} {\bibfnamefont {R.}~\bibnamefont {Rohs}}, \ and\ \bibinfo
  {author} {\bibfnamefont {D.~A.}\ \bibnamefont {Lidar}},\ }\href {\doibase
  10.1038/s41534-018-0060-8} {\bibfield  {journal} {\bibinfo  {journal} {npj
  Quantum Information}\ }\textbf {\bibinfo {volume} {4}},\ \bibinfo {pages}
  {14} (\bibinfo {year} {2018})}\BibitemShut {NoStop}%
\bibitem [{\citenamefont {{Venturelli}}\ \emph {et~al.}(2015)\citenamefont
  {{Venturelli}}, \citenamefont {{Marchand}},\ and\ \citenamefont
  {{Rojo}}}]{Venturelli2015}%
  \BibitemOpen
  \bibfield  {author} {\bibinfo {author} {\bibfnamefont {D.}~\bibnamefont
  {{Venturelli}}}, \bibinfo {author} {\bibfnamefont {D.~J.~J.}\ \bibnamefont
  {{Marchand}}}, \ and\ \bibinfo {author} {\bibfnamefont {G.}~\bibnamefont
  {{Rojo}}},\ }\href@noop {} {\bibfield  {journal} {\bibinfo  {journal} {ArXiv
  e-prints}\ } (\bibinfo {year} {2015})},\ \Eprint
  {http://arxiv.org/abs/1506.08479} {arXiv:1506.08479 [quant-ph]} \BibitemShut
  {NoStop}%
\bibitem [{\citenamefont {Neukart}\ \emph {et~al.}(2017)\citenamefont
  {Neukart}, \citenamefont {Compostella}, \citenamefont {Seidel}, \citenamefont
  {von Dollen}, \citenamefont {Yarkoni},\ and\ \citenamefont
  {Parney}}]{Neukart2017}%
  \BibitemOpen
  \bibfield  {author} {\bibinfo {author} {\bibfnamefont {F.}~\bibnamefont
  {Neukart}}, \bibinfo {author} {\bibfnamefont {G.}~\bibnamefont
  {Compostella}}, \bibinfo {author} {\bibfnamefont {C.}~\bibnamefont {Seidel}},
  \bibinfo {author} {\bibfnamefont {D.}~\bibnamefont {von Dollen}}, \bibinfo
  {author} {\bibfnamefont {S.}~\bibnamefont {Yarkoni}}, \ and\ \bibinfo
  {author} {\bibfnamefont {B.}~\bibnamefont {Parney}},\ }\href@noop {}
  {\bibfield  {journal} {\bibinfo  {journal} {Frontiers in ICT}\ }\textbf
  {\bibinfo {volume} {4}},\ \bibinfo {pages} {29} (\bibinfo {year}
  {2017})}\BibitemShut {NoStop}%
\bibitem [{\citenamefont {{Henderson}}\ \emph {et~al.}(2018)\citenamefont
  {{Henderson}}, \citenamefont {{Novak}},\ and\ \citenamefont
  {{Cook}}}]{Henderson2018}%
  \BibitemOpen
  \bibfield  {author} {\bibinfo {author} {\bibfnamefont {M.}~\bibnamefont
  {{Henderson}}}, \bibinfo {author} {\bibfnamefont {J.}~\bibnamefont
  {{Novak}}}, \ and\ \bibinfo {author} {\bibfnamefont {T.}~\bibnamefont
  {{Cook}}},\ }\href@noop {} {\bibfield  {journal} {\bibinfo  {journal} {ArXiv
  e-prints}\ } (\bibinfo {year} {2018})},\ \Eprint
  {http://arxiv.org/abs/1802.00069} {arXiv:1802.00069 [quant-ph]} \BibitemShut
  {NoStop}%
\bibitem [{\citenamefont {{Crawford}}\ \emph {et~al.}(2016)\citenamefont
  {{Crawford}}, \citenamefont {{Levit}}, \citenamefont {{Ghadermarzy}},
  \citenamefont {{Oberoi}},\ and\ \citenamefont {{Ronagh}}}]{Crawford2016}%
  \BibitemOpen
  \bibfield  {author} {\bibinfo {author} {\bibfnamefont {D.}~\bibnamefont
  {{Crawford}}}, \bibinfo {author} {\bibfnamefont {A.}~\bibnamefont {{Levit}}},
  \bibinfo {author} {\bibfnamefont {N.}~\bibnamefont {{Ghadermarzy}}}, \bibinfo
  {author} {\bibfnamefont {J.~S.}\ \bibnamefont {{Oberoi}}}, \ and\ \bibinfo
  {author} {\bibfnamefont {P.}~\bibnamefont {{Ronagh}}},\ }\href@noop {}
  {\bibfield  {journal} {\bibinfo  {journal} {ArXiv e-prints}\ } (\bibinfo
  {year} {2016})},\ \Eprint {http://arxiv.org/abs/1612.05695} {arXiv:1612.05695
  [quant-ph]} \BibitemShut {NoStop}%
\bibitem [{\citenamefont {Neukart}\ \emph {et~al.}(2018)\citenamefont
  {Neukart}, \citenamefont {Von~Dollen}, \citenamefont {Seidel},\ and\
  \citenamefont {Compostella}}]{Neukart2018}%
  \BibitemOpen
  \bibfield  {author} {\bibinfo {author} {\bibfnamefont {F.}~\bibnamefont
  {Neukart}}, \bibinfo {author} {\bibfnamefont {D.}~\bibnamefont {Von~Dollen}},
  \bibinfo {author} {\bibfnamefont {C.}~\bibnamefont {Seidel}}, \ and\ \bibinfo
  {author} {\bibfnamefont {G.}~\bibnamefont {Compostella}},\ }\href {\doibase
  10.3389/fphy.2017.00071} {\bibfield  {journal} {\bibinfo  {journal}
  {Frontiers in Physics}\ }\textbf {\bibinfo {volume} {5}},\ \bibinfo {pages}
  {71} (\bibinfo {year} {2018})}\BibitemShut {NoStop}%
\bibitem [{\citenamefont {Khoshaman}\ \emph {et~al.}(2018)\citenamefont
  {Khoshaman}, \citenamefont {Vinci}, \citenamefont {Denis}, \citenamefont
  {Andriyash},\ and\ \citenamefont {Amin}}]{Khoshaman2018}%
  \BibitemOpen
  \bibfield  {author} {\bibinfo {author} {\bibfnamefont {A.}~\bibnamefont
  {Khoshaman}}, \bibinfo {author} {\bibfnamefont {W.}~\bibnamefont {Vinci}},
  \bibinfo {author} {\bibfnamefont {B.}~\bibnamefont {Denis}}, \bibinfo
  {author} {\bibfnamefont {E.}~\bibnamefont {Andriyash}}, \ and\ \bibinfo
  {author} {\bibfnamefont {M.~H.}\ \bibnamefont {Amin}},\ }\href
  {http://stacks.iop.org/2058-9565/4/i=1/a=014001} {\bibfield  {journal}
  {\bibinfo  {journal} {Quantum Science and Technology}\ }\textbf {\bibinfo
  {volume} {4}},\ \bibinfo {pages} {014001} (\bibinfo {year}
  {2018})}\BibitemShut {NoStop}%
\bibitem [{\citenamefont {{Ohzeki}}\ \emph {et~al.}(2018)\citenamefont
  {{Ohzeki}}, \citenamefont {{Miki}}, \citenamefont {{Miyama}},\ and\
  \citenamefont {{Terabe}}}]{Ohzeki2019}%
  \BibitemOpen
  \bibfield  {author} {\bibinfo {author} {\bibfnamefont {M.}~\bibnamefont
  {{Ohzeki}}}, \bibinfo {author} {\bibfnamefont {A.}~\bibnamefont {{Miki}}},
  \bibinfo {author} {\bibfnamefont {M.~J.}\ \bibnamefont {{Miyama}}}, \ and\
  \bibinfo {author} {\bibfnamefont {M.}~\bibnamefont {{Terabe}}},\ }\href@noop
  {} {\bibfield  {journal} {\bibinfo  {journal} {ArXiv e-prints}\ } (\bibinfo
  {year} {2018})},\ \Eprint {http://arxiv.org/abs/1812.01532} {arXiv:1812.01532
  [quant-ph]} \BibitemShut {NoStop}%
\bibitem [{\citenamefont {Arai}\ \emph {et~al.}(2018)\citenamefont {Arai},
  \citenamefont {Ohzeki},\ and\ \citenamefont {Tanaka}}]{Arai2018nn}%
  \BibitemOpen
  \bibfield  {author} {\bibinfo {author} {\bibfnamefont {S.}~\bibnamefont
  {Arai}}, \bibinfo {author} {\bibfnamefont {M.}~\bibnamefont {Ohzeki}}, \ and\
  \bibinfo {author} {\bibfnamefont {K.}~\bibnamefont {Tanaka}},\ }\href
  {\doibase 10.7566/JPSJ.87.033001} {\bibfield  {journal} {\bibinfo  {journal}
  {Journal of the Physical Society of Japan}\ }\textbf {\bibinfo {volume}
  {87}},\ \bibinfo {pages} {033001} (\bibinfo {year} {2018})}\BibitemShut
  {NoStop}%
\bibitem [{\citenamefont {Takahashi}\ \emph {et~al.}(2018)\citenamefont
  {Takahashi}, \citenamefont {Ohzeki}, \citenamefont {Okada}, \citenamefont
  {Terabe}, \citenamefont {Taguchi},\ and\ \citenamefont
  {Tanaka}}]{Takahashi2018}%
  \BibitemOpen
  \bibfield  {author} {\bibinfo {author} {\bibfnamefont {C.}~\bibnamefont
  {Takahashi}}, \bibinfo {author} {\bibfnamefont {M.}~\bibnamefont {Ohzeki}},
  \bibinfo {author} {\bibfnamefont {S.}~\bibnamefont {Okada}}, \bibinfo
  {author} {\bibfnamefont {M.}~\bibnamefont {Terabe}}, \bibinfo {author}
  {\bibfnamefont {S.}~\bibnamefont {Taguchi}}, \ and\ \bibinfo {author}
  {\bibfnamefont {K.}~\bibnamefont {Tanaka}},\ }\href {\doibase
  10.7566/JPSJ.87.074001} {\bibfield  {journal} {\bibinfo  {journal} {Journal
  of the Physical Society of Japan}\ }\textbf {\bibinfo {volume} {87}},\
  \bibinfo {pages} {074001} (\bibinfo {year} {2018})}\BibitemShut {NoStop}%
\bibitem [{\citenamefont {Ohzeki}\ \emph {et~al.}(2018)\citenamefont {Ohzeki},
  \citenamefont {Takahashi}, \citenamefont {Okada}, \citenamefont {Terabe},
  \citenamefont {Taguchi},\ and\ \citenamefont {Tanaka}}]{Ohzeki2018NOLTA}%
  \BibitemOpen
  \bibfield  {author} {\bibinfo {author} {\bibfnamefont {M.}~\bibnamefont
  {Ohzeki}}, \bibinfo {author} {\bibfnamefont {C.}~\bibnamefont {Takahashi}},
  \bibinfo {author} {\bibfnamefont {S.}~\bibnamefont {Okada}}, \bibinfo
  {author} {\bibfnamefont {M.}~\bibnamefont {Terabe}}, \bibinfo {author}
  {\bibfnamefont {S.}~\bibnamefont {Taguchi}}, \ and\ \bibinfo {author}
  {\bibfnamefont {K.}~\bibnamefont {Tanaka}},\ }\href {\doibase
  10.1587/nolta.9.392} {\bibfield  {journal} {\bibinfo  {journal} {Nonlinear
  Theory and Its Applications, IEICE}\ }\textbf {\bibinfo {volume} {9}},\
  \bibinfo {pages} {392} (\bibinfo {year} {2018})}\BibitemShut {NoStop}%
\bibitem [{\citenamefont {Okada}\ \emph
  {et~al.}(2019{\natexlab{a}})\citenamefont {Okada}, \citenamefont {Ohzeki},
  \citenamefont {Terabe},\ and\ \citenamefont {Taguchi}}]{Okada2019}%
  \BibitemOpen
  \bibfield  {author} {\bibinfo {author} {\bibfnamefont {S.}~\bibnamefont
  {Okada}}, \bibinfo {author} {\bibfnamefont {M.}~\bibnamefont {Ohzeki}},
  \bibinfo {author} {\bibfnamefont {M.}~\bibnamefont {Terabe}}, \ and\ \bibinfo
  {author} {\bibfnamefont {S.}~\bibnamefont {Taguchi}},\ }\href@noop {}
  {\bibfield  {journal} {\bibinfo  {journal} {ArXiv e-prints}\ } (\bibinfo
  {year} {2019}{\natexlab{a}})},\ \Eprint {http://arxiv.org/abs/1901.00924}
  {arXiv:1901.00924 [quant-ph]} \BibitemShut {NoStop}%
\bibitem [{\citenamefont {Suzuki}\ and\ \citenamefont
  {Okada}(2005)}]{adiabatic_therem}%
  \BibitemOpen
  \bibfield  {author} {\bibinfo {author} {\bibfnamefont {S.}~\bibnamefont
  {Suzuki}}\ and\ \bibinfo {author} {\bibfnamefont {M.}~\bibnamefont {Okada}},\
  }\href@noop {} {\bibfield  {journal} {\bibinfo  {journal} {Journal of the
  Physical Society of Japan}\ }\textbf {\bibinfo {volume} {74}},\ \bibinfo
  {pages} {1649} (\bibinfo {year} {2005})}\BibitemShut {NoStop}%
\bibitem [{\citenamefont {Seki}\ and\ \citenamefont
  {Nishimori}(2012)}]{seki_antiferro}%
  \BibitemOpen
  \bibfield  {author} {\bibinfo {author} {\bibfnamefont {Y.}~\bibnamefont
  {Seki}}\ and\ \bibinfo {author} {\bibfnamefont {H.}~\bibnamefont
  {Nishimori}},\ }\href {\doibase 10.1103/PhysRevE.85.051112} {\bibfield
  {journal} {\bibinfo  {journal} {Phys. Rev. E}\ }\textbf {\bibinfo {volume}
  {85}},\ \bibinfo {pages} {051112} (\bibinfo {year} {2012})}\BibitemShut
  {NoStop}%
\bibitem [{\citenamefont {Seki}\ and\ \citenamefont
  {Nishimori}(2015)}]{seki_hop}%
  \BibitemOpen
  \bibfield  {author} {\bibinfo {author} {\bibfnamefont {Y.}~\bibnamefont
  {Seki}}\ and\ \bibinfo {author} {\bibfnamefont {H.}~\bibnamefont
  {Nishimori}},\ }\href {http://stacks.iop.org/1751-8121/48/i=33/a=335301}
  {\bibfield  {journal} {\bibinfo  {journal} {Journal of Physics A:
  Mathematical and Theoretical}\ }\textbf {\bibinfo {volume} {48}},\ \bibinfo
  {pages} {335301} (\bibinfo {year} {2015})}\BibitemShut {NoStop}%
\bibitem [{\citenamefont {Matsuura}\ \emph {et~al.}(2016)\citenamefont
  {Matsuura}, \citenamefont {Nishimori}, \citenamefont {Albash},\ and\
  \citenamefont {Lidar}}]{matsuura_prl}%
  \BibitemOpen
  \bibfield  {author} {\bibinfo {author} {\bibfnamefont {S.}~\bibnamefont
  {Matsuura}}, \bibinfo {author} {\bibfnamefont {H.}~\bibnamefont {Nishimori}},
  \bibinfo {author} {\bibfnamefont {T.}~\bibnamefont {Albash}}, \ and\ \bibinfo
  {author} {\bibfnamefont {D.~A.}\ \bibnamefont {Lidar}},\ }\href {\doibase
  10.1103/PhysRevLett.116.220501} {\bibfield  {journal} {\bibinfo  {journal}
  {Phys. Rev. Lett.}\ }\textbf {\bibinfo {volume} {116}},\ \bibinfo {pages}
  {220501} (\bibinfo {year} {2016})}\BibitemShut {NoStop}%
\bibitem [{\citenamefont {Okada}\ \emph
  {et~al.}(2019{\natexlab{b}})\citenamefont {Okada}, \citenamefont {Ohzeki},\
  and\ \citenamefont {Tanaka}}]{Okada_1dxx}%
  \BibitemOpen
  \bibfield  {author} {\bibinfo {author} {\bibfnamefont {S.}~\bibnamefont
  {Okada}}, \bibinfo {author} {\bibfnamefont {M.}~\bibnamefont {Ohzeki}}, \
  and\ \bibinfo {author} {\bibfnamefont {K.}~\bibnamefont {Tanaka}},\ }\href
  {\doibase 10.7566/JPSJ.88.024802} {\bibfield  {journal} {\bibinfo  {journal}
  {Journal of the Physical Society of Japan}\ }\textbf {\bibinfo {volume}
  {88}},\ \bibinfo {pages} {024802} (\bibinfo {year}
  {2019}{\natexlab{b}})}\BibitemShut {NoStop}%
\bibitem [{\citenamefont {Suzuki}(1976)}]{suzu_toro}%
  \BibitemOpen
  \bibfield  {author} {\bibinfo {author} {\bibfnamefont {M.}~\bibnamefont
  {Suzuki}},\ }\href {\doibase 10.1007/BF01609348} {\bibfield  {journal}
  {\bibinfo  {journal} {Communications in Mathematical Physics}\ }\textbf
  {\bibinfo {volume} {51}},\ \bibinfo {pages} {183} (\bibinfo {year}
  {1976})}\BibitemShut {NoStop}%
\bibitem [{\citenamefont {Bravyi}\ \emph {et~al.}(2008)\citenamefont {Bravyi},
  \citenamefont {Divincenzo}, \citenamefont {Oliveira},\ and\ \citenamefont
  {Terhal}}]{Bravyi2008}%
  \BibitemOpen
  \bibfield  {author} {\bibinfo {author} {\bibfnamefont {S.}~\bibnamefont
  {Bravyi}}, \bibinfo {author} {\bibfnamefont {D.~P.}\ \bibnamefont
  {Divincenzo}}, \bibinfo {author} {\bibfnamefont {R.}~\bibnamefont
  {Oliveira}}, \ and\ \bibinfo {author} {\bibfnamefont {B.~M.}\ \bibnamefont
  {Terhal}},\ }\href@noop {} {\bibfield  {journal} {\bibinfo  {journal}
  {Quantum Info. Comput.}\ }\textbf {\bibinfo {volume} {8}},\ \bibinfo {pages}
  {361} (\bibinfo {year} {2008})}\BibitemShut {NoStop}%
\bibitem [{\citenamefont {Hormozi}\ \emph {et~al.}(2017)\citenamefont
  {Hormozi}, \citenamefont {Brown}, \citenamefont {Carleo},\ and\ \citenamefont
  {Troyer}}]{non_stoquastic}%
  \BibitemOpen
  \bibfield  {author} {\bibinfo {author} {\bibfnamefont {L.}~\bibnamefont
  {Hormozi}}, \bibinfo {author} {\bibfnamefont {E.~W.}\ \bibnamefont {Brown}},
  \bibinfo {author} {\bibfnamefont {G.}~\bibnamefont {Carleo}}, \ and\ \bibinfo
  {author} {\bibfnamefont {M.}~\bibnamefont {Troyer}},\ }\href {\doibase
  10.1103/PhysRevB.95.184416} {\bibfield  {journal} {\bibinfo  {journal} {Phys.
  Rev. B}\ }\textbf {\bibinfo {volume} {95}},\ \bibinfo {pages} {184416}
  (\bibinfo {year} {2017})}\BibitemShut {NoStop}%
\bibitem [{\citenamefont {Nishimori}\ and\ \citenamefont
  {Takada}(2017)}]{expo_enhancement}%
  \BibitemOpen
  \bibfield  {author} {\bibinfo {author} {\bibfnamefont {H.}~\bibnamefont
  {Nishimori}}\ and\ \bibinfo {author} {\bibfnamefont {K.}~\bibnamefont
  {Takada}},\ }\href {\doibase 10.3389/fict.2017.00002} {\bibfield  {journal}
  {\bibinfo  {journal} {Frontiers in ICT}\ }\textbf {\bibinfo {volume} {4}},\
  \bibinfo {pages} {2} (\bibinfo {year} {2017})}\BibitemShut {NoStop}%
\bibitem [{\citenamefont {Ohzeki}(2017)}]{adqmc}%
  \BibitemOpen
  \bibfield  {author} {\bibinfo {author} {\bibfnamefont {M.}~\bibnamefont
  {Ohzeki}},\ }\href {http://dx.doi.org/10.1038/srep41186} {\bibfield
  {journal} {\bibinfo  {journal} {Scientific Reports}\ }\textbf {\bibinfo
  {volume} {7}},\ \bibinfo {pages} {41186 EP } (\bibinfo {year}
  {2017})}\BibitemShut {NoStop}%
\bibitem [{\citenamefont {Evgeny}\ and\ \citenamefont
  {Amin}(2017)}]{can_qmc_simulate_QA}%
  \BibitemOpen
  \bibfield  {author} {\bibinfo {author} {\bibfnamefont {A.}~\bibnamefont
  {Evgeny}}\ and\ \bibinfo {author} {\bibfnamefont {M.~H.}\ \bibnamefont
  {Amin}},\ }\href@noop {} {\bibfield  {journal} {\bibinfo  {journal} {ArXiv
  e-prints}\ } (\bibinfo {year} {2017})},\ \Eprint
  {http://arxiv.org/abs/1703.09277} {arXiv:1703.09277 [quant-ph]} \BibitemShut
  {NoStop}%
\bibitem [{\citenamefont {Isakov}\ \emph {et~al.}(2016)\citenamefont {Isakov},
  \citenamefont {Mazzola}, \citenamefont {Smelyanskiy}, \citenamefont {Jiang},
  \citenamefont {Boixo}, \citenamefont {Neven},\ and\ \citenamefont
  {Troyer}}]{qmc_tunneling}%
  \BibitemOpen
  \bibfield  {author} {\bibinfo {author} {\bibfnamefont {S.~V.}\ \bibnamefont
  {Isakov}}, \bibinfo {author} {\bibfnamefont {G.}~\bibnamefont {Mazzola}},
  \bibinfo {author} {\bibfnamefont {V.~N.}\ \bibnamefont {Smelyanskiy}},
  \bibinfo {author} {\bibfnamefont {Z.}~\bibnamefont {Jiang}}, \bibinfo
  {author} {\bibfnamefont {S.}~\bibnamefont {Boixo}}, \bibinfo {author}
  {\bibfnamefont {H.}~\bibnamefont {Neven}}, \ and\ \bibinfo {author}
  {\bibfnamefont {M.}~\bibnamefont {Troyer}},\ }\href {\doibase
  10.1103/PhysRevLett.117.180402} {\bibfield  {journal} {\bibinfo  {journal}
  {Phys. Rev. Lett.}\ }\textbf {\bibinfo {volume} {117}},\ \bibinfo {pages}
  {180402} (\bibinfo {year} {2016})}\BibitemShut {NoStop}%
\bibitem [{\citenamefont {Jiang}\ \emph {et~al.}(2017)\citenamefont {Jiang},
  \citenamefont {Smelyanskiy}, \citenamefont {Isakov}, \citenamefont {Boixo},
  \citenamefont {Mazzola}, \citenamefont {Troyer},\ and\ \citenamefont
  {Neven}}]{scaling_qmc}%
  \BibitemOpen
  \bibfield  {author} {\bibinfo {author} {\bibfnamefont {Z.}~\bibnamefont
  {Jiang}}, \bibinfo {author} {\bibfnamefont {V.~N.}\ \bibnamefont
  {Smelyanskiy}}, \bibinfo {author} {\bibfnamefont {S.~V.}\ \bibnamefont
  {Isakov}}, \bibinfo {author} {\bibfnamefont {S.}~\bibnamefont {Boixo}},
  \bibinfo {author} {\bibfnamefont {G.}~\bibnamefont {Mazzola}}, \bibinfo
  {author} {\bibfnamefont {M.}~\bibnamefont {Troyer}}, \ and\ \bibinfo {author}
  {\bibfnamefont {H.}~\bibnamefont {Neven}},\ }\href {\doibase
  10.1103/PhysRevA.95.012322} {\bibfield  {journal} {\bibinfo  {journal} {Phys.
  Rev. A}\ }\textbf {\bibinfo {volume} {95}},\ \bibinfo {pages} {012322}
  (\bibinfo {year} {2017})}\BibitemShut {NoStop}%
\bibitem [{\citenamefont {Kirkpatrick}\ and\ \citenamefont
  {Thirumalai}(1987)}]{p_spin_kirk}%
  \BibitemOpen
  \bibfield  {author} {\bibinfo {author} {\bibfnamefont {T.~R.}\ \bibnamefont
  {Kirkpatrick}}\ and\ \bibinfo {author} {\bibfnamefont {D.}~\bibnamefont
  {Thirumalai}},\ }\href {\doibase 10.1103/PhysRevB.36.5388} {\bibfield
  {journal} {\bibinfo  {journal} {Phys. Rev. B}\ }\textbf {\bibinfo {volume}
  {36}},\ \bibinfo {pages} {5388} (\bibinfo {year} {1987})}\BibitemShut
  {NoStop}%
\bibitem [{\citenamefont {Swift}\ \emph {et~al.}(2000)\citenamefont {Swift},
  \citenamefont {Bokil}, \citenamefont {Travasso},\ and\ \citenamefont
  {Bray}}]{glass_prb}%
  \BibitemOpen
  \bibfield  {author} {\bibinfo {author} {\bibfnamefont {M.~R.}\ \bibnamefont
  {Swift}}, \bibinfo {author} {\bibfnamefont {H.}~\bibnamefont {Bokil}},
  \bibinfo {author} {\bibfnamefont {R.~D.~M.}\ \bibnamefont {Travasso}}, \ and\
  \bibinfo {author} {\bibfnamefont {A.~J.}\ \bibnamefont {Bray}},\ }\href
  {\doibase 10.1103/PhysRevB.62.11494} {\bibfield  {journal} {\bibinfo
  {journal} {Phys. Rev. B}\ }\textbf {\bibinfo {volume} {62}},\ \bibinfo
  {pages} {11494} (\bibinfo {year} {2000})}\BibitemShut {NoStop}%
\bibitem [{\citenamefont {Seoane}\ and\ \citenamefont
  {Nishimori}(2012)}]{Seoane}%
  \BibitemOpen
  \bibfield  {author} {\bibinfo {author} {\bibfnamefont {B.}~\bibnamefont
  {Seoane}}\ and\ \bibinfo {author} {\bibfnamefont {H.}~\bibnamefont
  {Nishimori}},\ }\href {http://stacks.iop.org/1751-8121/45/i=43/a=435301}
  {\bibfield  {journal} {\bibinfo  {journal} {Journal of Physics A:
  Mathematical and Theoretical}\ }\textbf {\bibinfo {volume} {45}},\ \bibinfo
  {pages} {435301} (\bibinfo {year} {2012})}\BibitemShut {NoStop}%
\bibitem [{\citenamefont {Matsuura}\ \emph {et~al.}(2017)\citenamefont
  {Matsuura}, \citenamefont {Nishimori}, \citenamefont {Vinci}, \citenamefont
  {Albash},\ and\ \citenamefont {Lidar}}]{QAC}%
  \BibitemOpen
  \bibfield  {author} {\bibinfo {author} {\bibfnamefont {S.}~\bibnamefont
  {Matsuura}}, \bibinfo {author} {\bibfnamefont {H.}~\bibnamefont {Nishimori}},
  \bibinfo {author} {\bibfnamefont {W.}~\bibnamefont {Vinci}}, \bibinfo
  {author} {\bibfnamefont {T.}~\bibnamefont {Albash}}, \ and\ \bibinfo {author}
  {\bibfnamefont {D.~A.}\ \bibnamefont {Lidar}},\ }\href {\doibase
  10.1103/PhysRevA.95.022308} {\bibfield  {journal} {\bibinfo  {journal} {Phys.
  Rev. A}\ }\textbf {\bibinfo {volume} {95}},\ \bibinfo {pages} {022308}
  (\bibinfo {year} {2017})}\BibitemShut {NoStop}%
\bibitem [{\citenamefont {Bertalan}\ \emph {et~al.}(2011)\citenamefont
  {Bertalan}, \citenamefont {Kuma}, \citenamefont {Matsuda},\ and\
  \citenamefont {Nishimori}}]{ensemble_pspin}%
  \BibitemOpen
  \bibfield  {author} {\bibinfo {author} {\bibfnamefont {Z.}~\bibnamefont
  {Bertalan}}, \bibinfo {author} {\bibfnamefont {T.}~\bibnamefont {Kuma}},
  \bibinfo {author} {\bibfnamefont {Y.}~\bibnamefont {Matsuda}}, \ and\
  \bibinfo {author} {\bibfnamefont {H.}~\bibnamefont {Nishimori}},\ }\href
  {http://stacks.iop.org/1742-5468/2011/i=01/a=P01016} {\bibfield  {journal}
  {\bibinfo  {journal} {Journal of Statistical Mechanics: Theory and
  Experiment}\ }\textbf {\bibinfo {volume} {2011}},\ \bibinfo {pages} {P01016}
  (\bibinfo {year} {2011})}\BibitemShut {NoStop}%
\bibitem [{\citenamefont {Inoue}\ and\ \citenamefont
  {Tanaka}(2001)}]{dy_maxmum_like}%
  \BibitemOpen
  \bibfield  {author} {\bibinfo {author} {\bibfnamefont {J.}~\bibnamefont
  {Inoue}}\ and\ \bibinfo {author} {\bibfnamefont {K.}~\bibnamefont {Tanaka}},\
  }\href {\doibase 10.1103/PhysRevE.65.016125} {\bibfield  {journal} {\bibinfo
  {journal} {Phys. Rev. E}\ }\textbf {\bibinfo {volume} {65}},\ \bibinfo
  {pages} {016125} (\bibinfo {year} {2001})}\BibitemShut {NoStop}%
\bibitem [{\citenamefont {Inoue}(2010)}]{determ_order_qmc}%
  \BibitemOpen
  \bibfield  {author} {\bibinfo {author} {\bibfnamefont {J.}~\bibnamefont
  {Inoue}},\ }\href@noop {} {\bibfield  {journal} {\bibinfo  {journal} {J.
  Phys.Conf. Ser.}\ }\textbf {\bibinfo {volume} {233}},\ \bibinfo {pages}
  {012010} (\bibinfo {year} {2010})}\BibitemShut {NoStop}%
\bibitem [{\citenamefont {Inoue}(2011)}]{inoue_hop}%
  \BibitemOpen
  \bibfield  {author} {\bibinfo {author} {\bibfnamefont {J.}~\bibnamefont
  {Inoue}},\ }\href {http://stacks.iop.org/1742-6596/297/i=1/a=012012}
  {\bibfield  {journal} {\bibinfo  {journal} {J. Phys.Conf. Ser.}\ }\textbf
  {\bibinfo {volume} {297}},\ \bibinfo {pages} {012012} (\bibinfo {year}
  {2011})}\BibitemShut {NoStop}%
\bibitem [{\citenamefont {Strogatz}(2000)}]{Strogatz}%
  \BibitemOpen
  \bibfield  {author} {\bibinfo {author} {\bibfnamefont {S.~H.}\ \bibnamefont
  {Strogatz}},\ }\href@noop {} {\emph {\bibinfo {title} {Nonlinear Dynamics and
  Chaos: With Applications to Physics, Biology, Chemistry and Engineering}}}\
  (\bibinfo  {publisher} {Westview Press},\ \bibinfo {year} {2000})\BibitemShut
  {NoStop}%
\bibitem [{\citenamefont {Liu}\ and\ \citenamefont
  {Elaydi}(2001)}]{stability_1}%
  \BibitemOpen
  \bibfield  {author} {\bibinfo {author} {\bibfnamefont {P.}~\bibnamefont
  {Liu}}\ and\ \bibinfo {author} {\bibfnamefont {S.~N.}\ \bibnamefont
  {Elaydi}},\ }\href {\doibase 10.1023/A:1011539901001} {\bibfield  {journal}
  {\bibinfo  {journal} {Journal of Computational Analysis and Applications}\
  }\textbf {\bibinfo {volume} {3}},\ \bibinfo {pages} {53} (\bibinfo {year}
  {2001})}\BibitemShut {NoStop}%
\bibitem [{\citenamefont {Supajaidee}\ and\ \citenamefont
  {Moonchai}(2017)}]{stability_2}%
  \BibitemOpen
  \bibfield  {author} {\bibinfo {author} {\bibfnamefont {N.}~\bibnamefont
  {Supajaidee}}\ and\ \bibinfo {author} {\bibfnamefont {S.}~\bibnamefont
  {Moonchai}},\ }\href {\doibase 10.1186/s13662-017-1430-9} {\bibfield
  {journal} {\bibinfo  {journal} {Advances in Difference Equations}\ }\textbf
  {\bibinfo {volume} {2017}},\ \bibinfo {pages} {372} (\bibinfo {year}
  {2017})}\BibitemShut {NoStop}%
\bibitem [{\citenamefont {J{\"o}rg}\ \emph {et~al.}(2010)\citenamefont
  {J{\"o}rg}, \citenamefont {Krzakala}, \citenamefont {Kurchan}, \citenamefont
  {Maggs},\ and\ \citenamefont {Pujos}}]{p_ising}%
  \BibitemOpen
  \bibfield  {author} {\bibinfo {author} {\bibfnamefont {T.}~\bibnamefont
  {J{\"o}rg}}, \bibinfo {author} {\bibfnamefont {F.}~\bibnamefont {Krzakala}},
  \bibinfo {author} {\bibfnamefont {J.}~\bibnamefont {Kurchan}}, \bibinfo
  {author} {\bibfnamefont {A.~C.}\ \bibnamefont {Maggs}}, \ and\ \bibinfo
  {author} {\bibfnamefont {J.}~\bibnamefont {Pujos}},\ }\href
  {http://stacks.iop.org/0295-5075/89/i=4/a=40004} {\bibfield  {journal}
  {\bibinfo  {journal} {Europhys. Lett.}\ }\textbf {\bibinfo {volume} {89}},\
  \bibinfo {pages} {40004} (\bibinfo {year} {2010})}\BibitemShut {NoStop}%
\bibitem [{\citenamefont {Chancellor}(2017)}]{reverse_anneal}%
  \BibitemOpen
  \bibfield  {author} {\bibinfo {author} {\bibfnamefont {N.}~\bibnamefont
  {Chancellor}},\ }\href {http://stacks.iop.org/1367-2630/19/i=2/a=023024}
  {\bibfield  {journal} {\bibinfo  {journal} {New Journal of Physics}\ }\textbf
  {\bibinfo {volume} {19}},\ \bibinfo {pages} {023024} (\bibinfo {year}
  {2017})}\BibitemShut {NoStop}%
\bibitem [{\citenamefont {Susa}\ \emph
  {et~al.}(2018{\natexlab{a}})\citenamefont {Susa}, \citenamefont {Yamashiro},
  \citenamefont {Yamamoto},\ and\ \citenamefont {Nishimori}}]{inhomo}%
  \BibitemOpen
  \bibfield  {author} {\bibinfo {author} {\bibfnamefont {Y.}~\bibnamefont
  {Susa}}, \bibinfo {author} {\bibfnamefont {Y.}~\bibnamefont {Yamashiro}},
  \bibinfo {author} {\bibfnamefont {M.}~\bibnamefont {Yamamoto}}, \ and\
  \bibinfo {author} {\bibfnamefont {H.}~\bibnamefont {Nishimori}},\ }\href@noop
  {} {\bibfield  {journal} {\bibinfo  {journal} {Journal of the Physical
  Society of Japan}\ }\textbf {\bibinfo {volume} {87}},\ \bibinfo {pages}
  {023002} (\bibinfo {year} {2018}{\natexlab{a}})}\BibitemShut {NoStop}%
\bibitem [{\citenamefont {Ohkuwa}\ \emph {et~al.}(2018)\citenamefont {Ohkuwa},
  \citenamefont {Nishimori},\ and\ \citenamefont {Lidar}}]{RA_ohkuwa}%
  \BibitemOpen
  \bibfield  {author} {\bibinfo {author} {\bibfnamefont {M.}~\bibnamefont
  {Ohkuwa}}, \bibinfo {author} {\bibfnamefont {H.}~\bibnamefont {Nishimori}}, \
  and\ \bibinfo {author} {\bibfnamefont {D.~A.}\ \bibnamefont {Lidar}},\ }\href
  {\doibase 10.1103/PhysRevA.98.022314} {\bibfield  {journal} {\bibinfo
  {journal} {Phys. Rev. A}\ }\textbf {\bibinfo {volume} {98}},\ \bibinfo
  {pages} {022314} (\bibinfo {year} {2018})}\BibitemShut {NoStop}%
\bibitem [{\citenamefont {Susa}\ \emph
  {et~al.}(2018{\natexlab{b}})\citenamefont {Susa}, \citenamefont {Yamashiro},
  \citenamefont {Yamamoto}, \citenamefont {Hen}, \citenamefont {Lidar},\ and\
  \citenamefont {Nishimori}}]{susa_inhomo}%
  \BibitemOpen
  \bibfield  {author} {\bibinfo {author} {\bibfnamefont {Y.}~\bibnamefont
  {Susa}}, \bibinfo {author} {\bibfnamefont {Y.}~\bibnamefont {Yamashiro}},
  \bibinfo {author} {\bibfnamefont {M.}~\bibnamefont {Yamamoto}}, \bibinfo
  {author} {\bibfnamefont {I.}~\bibnamefont {Hen}}, \bibinfo {author}
  {\bibfnamefont {D.~A.}\ \bibnamefont {Lidar}}, \ and\ \bibinfo {author}
  {\bibfnamefont {H.}~\bibnamefont {Nishimori}},\ }\href {\doibase
  10.1103/PhysRevA.98.042326} {\bibfield  {journal} {\bibinfo  {journal} {Phys.
  Rev. A}\ }\textbf {\bibinfo {volume} {98}},\ \bibinfo {pages} {042326}
  (\bibinfo {year} {2018}{\natexlab{b}})}\BibitemShut {NoStop}%
\end{thebibliography}%
\end{document}